\documentclass[aps,singlecolumn,pra,superscriptaddress,footinbib,notitlepage]{revtex4-1}
\usepackage{amsmath,amssymb,bm}
\usepackage{graphicx}
\usepackage{epstopdf}
\usepackage{latexsym}
\usepackage[usenames,dvipsnames]{color}
\usepackage{hyperref}
\usepackage{natbib}
\begin{document}

% \end{document}

% \usepackage[T1]{fontenc}
% % \usepackage{setspace}
% \usepackage[latin9]{inputenc}
% \setcounter{secnumdepth}{3}
% \usepackage{amsmath}
% \usepackage{amssymb}
% \usepackage{graphicx}
% \usepackage{hyperref}
% \usepackage{natbib}
% %\usepackage{esint}
% \usepackage{times}
% \usepackage{helvet}
% \usepackage{braket}
% \usepackage{babel}

% \renewcommand{\baselinestretch}{1.0}

\makeatletter
%%%%%%%%%%%%%%%%%%%%%%%%%%%%%% Textclass specific LaTeX commands.
 % Fix a bug in REVTeX 4.1
 \def\lovname{List of Videos}
 \@ifundefined{textcolor}{}
 {%
   \definecolor{BLACK}{gray}{0}
   \definecolor{WHITE}{gray}{1}
   \definecolor{RED}{rgb}{1,0,0}
   \definecolor{GREEN}{rgb}{0,1,0}
   \definecolor{BLUE}{rgb}{0,0,1}
   \definecolor{CYAN}{cmyk}{1,0,0,0}
   \definecolor{MAGENTA}{cmyk}{0,1,0,0}
   \definecolor{YELLOW}{cmyk}{0,0,1,0}
 }

%%%%%%%%%%%%%%%%%%%%%%%%%%%%%% User specified LaTeX commands.

%\usepackage{times}

\newcommand{\Rev }[1]{{\color{black}{#1}\normalcolor}} % Revision
\newcommand{\Com}[1]{{\color{black}{#1}\normalcolor}} %Comment
\newcommand{\LGPE}{\mathcal{L}}

%%%%%%%%%%%%%%%%%%%%%%%%%%%%%% Textclass specific LaTeX commands.
\@ifundefined{textcolor}{}{%
 \definecolor{BLACK}{gray}{0}
 \definecolor{WHITE}{gray}{1}
 \definecolor{RED}{rgb}{1,0,0}
 \definecolor{GREEN}{rgb}{0,1,0}
 \definecolor{BLUE}{rgb}{0,0,1}
 \definecolor{CYAN}{cmyk}{1,0,0,0}
 \definecolor{MAGENTA}{cmyk}{0,1,0,0}
 \definecolor{YELLOW}{cmyk}{0,0,1,0}
 }
\newcommand{\mjd}[1]{{\textbf{\textcolor{blue}{#1}}}}
\newcommand{\js}[1]{{\textbf{\textcolor{red}{#1}}}}
\newcommand{\ket}[1]{\ensuremath{\left|  #1 \right\rangle}}
\newcommand{\bra}[1]{\ensuremath{\left\langle  #1 \right|}}
\DeclareRobustCommand{\rchi}{{\mathpalette\irchi\relax}}
\newcommand{\irchi}[2]{\raisebox{\depth}{$#1\chi$}}

\@ifundefined{definecolor}{\@ifundefined{definecolor}
 {\@ifundefined{definecolor}
 {\@ifundefined{definecolor}
 {\@ifundefined{definecolor}
 {\@ifundefined{definecolor}
 {\usepackage{color}}{}
}{}
}{}
}{}
}{}
}{}
\setcounter{MaxMatrixCols}{10}

\makeatother

\title{
Unifying scrambling, thermalization and entanglement through the measurement of fidelity out-of-time-order correlators in the Dicke model
}

% \author{R.~J. Lewis-Swan$^{1,2,4}$, A. Safavi-Naini$^{1,2,4}$, J.~J. Bollinger$^{3}$, and A.~M. Rey$^{1,2}$}

\author{ R. J. Lewis-Swan}
\thanks{These two authors contributed equally}
\affiliation{JILA, NIST and University of Colorado, 440 UCB, Boulder, CO 80309, USA}
\affiliation{Center for Theory of Quantum Matter, University of Colorado, Boulder, CO 80309, USA}
\author{ A. Safavi-Naini}
\thanks{These two authors contributed equally}
\affiliation{JILA, NIST and University of Colorado, 440 UCB, Boulder, CO 80309, USA}
\affiliation{Center for Theory of Quantum Matter, University of Colorado, Boulder, CO 80309, USA}
\author{J. J. Bollinger}
\affiliation{NIST, Boulder, CO 80305, USA}
\author{A. M. Rey}
\affiliation{JILA, NIST and University of Colorado, 440 UCB, Boulder, CO 80309, USA}
\affiliation{Center for Theory of Quantum Matter, University of Colorado, Boulder, CO 80309, USA}

% 
% \begin{affiliations}
% \item JILA, NIST and Department of Physics, University of Colorado, Boulder, USA
% \item Center for Theory of Quantum Matter, University of Colorado, Boulder, CO 80309, USA
% \item NIST, Boulder, CO 80305, USA
% \item These authors contributed equally to this manuscript
% \end{affiliations}

%Currently 285 words -- limit 150
\begin{abstract}

Scrambling  is the process by which information stored in local degrees of freedom  spreads over the  many-body degrees of freedom of a quantum systems, becoming inaccessible to local probes and apparently lost. Scrambling and entanglement can  reconcile seemingly unrelated behaviors including thermalization of isolated quantum systems and information loss in black holes. Here, we demonstrate that fidelity out-of-time-order correlators (FOTOCs) can elucidate connections between  scrambling, entanglement, ergodicity and quantum chaos (butterfly effect). We compute FOTOCs for the paradigmatic Dicke model, and show they can  measure subsystem R\'{e}nyi entropies and inform about quantum thermalization. Moreover, we illustrate why  FOTOCs  give access to a simple relation between quantum and classical Lyapunov exponents in a chaotic system without finite-size effects. Our results open a path to experimental use  FOTOCs to explore scrambling, bounds on quantum information processing and investigate black hole analogs in controllable quantum systems.

\end{abstract}

\maketitle

\section{Introduction}

Recent studies have shown that isolated many-body quantum systems, under unitary time evolution, can become highly entangled and thus thermalize. This understanding has led to insights as to how statistical mechanics emerges in closed quantum systems
\cite{DAlessio2016,Nandkishore2015,Eisert2016}. Moreover, the relevance of entanglement as a resource for quantum information processing, quantum communication and metrology has stimulated cross-disciplinary efforts to  quantify and characterize entanglement. Experimental progress in controlling clean, highly isolated, and fully tunable quantum systems, where entanglement can be measured, have resulted in radical advances in this direction.  However, such measurements have  been restricted to few body systems, including arrays of $6 \times 2$ bosonic atoms \cite{Kaufman2016}, three superconducting qubits \cite{Martinis2016}, and systems of $\lesssim 20$ trapped ions \cite{Schaetz2016,Roos2018}. The model we study here and the measurements we propose can be implemented in trapped ions with more than 100 spins.

Concurrently, out-of-time-order correlations (OTOCs)  \cite{Hayden2007,Sekino2008,Shenker2014, Yoshida2016,Kitaev2015,Swingle2016}
\begin{equation}
    F(t)= \langle \hat{W}^\dagger(t) \hat{V}^\dagger \hat{W}(t) \hat{V}\rangle,
\label{eq:OTCF}
\end{equation}
have been identified as measures of the dynamics of quantum information scrambling. Here, $\hat{W}(t)=e^{i \hat{H} t}\hat{W} e^{-i \hat{H} t}$, with $\hat{H}$ a quantum many-body Hamiltonian, and $\hat{W}$ and $\hat{V}$ two initially commuting and unitary operators. Whilst OTOCs can be computed with respect to any (possibly mixed) state, here we focus on the case where the initial state of the system is pure. The quantity ${\rm Re}[F(t)] = 1-\langle [ \hat{V}^{\dagger}, \hat{W}^{\dagger}(t) ] [ \hat{W}(t), \hat{V}] \rangle/2$  \footnote{If {${\hat{V}}$} is not unitary but a projector, e.g.  ${\hat{V}\hat{V}^\dagger=\hat{V}}$ and ${ \hat{V} }$ commutes with the density matrix of the initial state then ${ {\rm Re}[F(t)] = 1-\langle [ \hat{V}^{\dagger}, \hat{W}^{\dagger}(t) ] [ \hat{W}(t), \hat{V}] \rangle }$ } encapsulates the degree that $\hat{W}(t)$ and $\hat{V}$ fail to commute at later times due to the time evolution of $\hat{W}$ under $\hat{H}$. The fastest scramblers \cite{Hayden2007,Sekino2008,Shenker2014,Maldacena2016}, such as black holes, feature an exponential growth of scrambling which manifests as $1-{\rm Re}[F(t)]\sim e^{\lambda_{\mathrm{Q}} t}$. Here, $\lambda_{\mathrm{Q}}$ is the quantum Lyapunov exponent which serves as a proxy for quantum chaos. Regardless of the OTOCs' apparent complexity \cite{Swingle2016,Yao2016,Shen2016,Zhu2016}, the capability to perform many-body echoes (see Fig.~\ref{fig:SchemFig}) in current experiments \cite{Garttner2016,GarttnerMQC2017,Li2017,Meier_2017} has opened a path for the experimental investigation of quantum scrambling, however
so far those  have  not probed quantum chaos or fast scrambling.

\Rev{
Here we show that FOTOCs, a specific family of fidelity out-of-time-order correlators, which set  $\hat{V}$  to be a projector on the initial state,  can provide profound insight on scrambling behavior. We explicitly compute FOTOCs in the in the Dicke model \cite{Dicke1954}, an iconic model in quantum optics, and  illustrate how FOTOCs elucidate theoretical connections between scrambling, volume-law R\'enyi entropy (RE) and thermalization, while linking quantum and classical chaos (Fig.\ref{fig:SchemFig}a). Additionally, we discuss how one can probe these connections readily in experiments.
}

\section{Results}

\subsection{Model}
\Rev{
The Dicke model (DM)\cite{Dicke1954}, describes the coupling a single large spin and a harmonic oscillator and has been recently implemented in atomic \cite{Baumann2010,Baumann2011,Klinder2015,Parkins2018} and trapped ion setups \cite{Dickeionpaper2018}.   The Hamiltonian of the DM is given by}
\begin{equation}
\label{eq:Hdicke}
 \hat H_{\rm D}= \frac{ 2 g}{\sqrt N} \left(\hat a + \hat a^\dagger\right) \hat S_z + \delta \hat a^\dag \hat a +  B \hat S_x,
\end{equation} where $B$ characterizes the strength of the transverse field, $\delta$ the detuning of the bosonic mode from the driving field with strength $g$ that generates the spin-boson coupling. Here, $g,\delta, B\geq0 $. The operator  $\hat{a}$ ($\hat{a}^{\dagger}$) is the bosonic annihilation (creation) operator of the mode, and $\hat{S}_{\alpha} = \sum_{j=1}^{N} \hat{\sigma}^{\alpha}_j/2$ are collective spin operators with $\hat{\sigma}^{\alpha}_j$ ($\alpha=x,y,z$) the Pauli matrices for the $j$th spin-$1/2$.

\subsection{Connections between scrambling dynamics and chaos}
Even when restricted to the Dicke manifold, i.e. states with $S=N/2$, with $S(S+1)$ the eigenvalue of the total spin operator $\hat{S}^2= \hat{S}_x^2+\hat{S}_y^2+\hat{S}_z^2$, this model exhibits rich physics (see Fig.~\ref{fig:phase}a). At zero temperature, $T=0$, the DM features a quantum phase transition (QPT) as the system crosses a critical field $B_{\mathrm{c}}=4 g^2/\delta$. For $B>B_{\mathrm{c}}$ (normal phase) the ground-state is described by spins aligned along the transverse field and a bosonic vacuum. For $B<B_{\mathrm{c}}$ (superradiant phase), the ground-state is ferromagnetic, $\langle |\hat{S}_Z|\rangle\sim N/2$, and characterized by macroscopic occupation of the bosonic mode  (Fig. \ref{fig:phase} a).
Furthermore, in the superradiant phase ($B<B_{\mathrm{c}}$), the DM features a family of excited-state quantum phase transitions (ESQPTs). The ESQPTs are signaled by singularities in the energy level structure  and  a change in the spectral statistics \cite{PerezFernandez2011,Brandes2013,Emary2003_PRE,Gritsev2017} at a critical energy $E_{\mathrm{c}}= -B N/2$ which coincides with the ground-state energy of the normal phase. Figure~\ref{fig:phase}a shows how the  nearest-neighbor spacing distribution $P(s)$, where $s$ is a normalized distance between two neighboring energy levels, features a different character on either side of $E_{\mathrm{c}}$. For $E>E_{\mathrm{c}}$ the spectral statistics are similar to the Wigner-Dyson distribution $P_W(s)=\pi s/2 \exp(-\pi s^2/4)$, which in random-matrix theory describes a chaotic system.
For $E<E_{\mathrm{c}}$ the shape of the histograms is neither  Wigner-Dyson nor Poissonian $P_P(s)=\exp(-s)$. The latter characterizes level statistics of non-ergodic systems, and is observed in the normal phase. Whilst the deviations from clear Wigner-Dyson or Poissonian statistics in regimes II and III is attributable to finite size effects \cite{PerezFernandez2011}, we emphasize that they clearly show a stark contrast in the degree of level-repulsion, which is a qualitative signature of quantum chaos.

Similar features appear in the classical dynamics of the DM \cite{Altland2012,Altland2012_NJP,Emary2003_PRL,Emary2003_PRE,ChavezCarlos2016}, manifested in the different behavior of  trajectories in phase-space computed from the mean-field equations of motion for: $\mathbf{\vec x}=(\langle \hat S_x\rangle , \langle \hat S_y \rangle , \langle \hat S_z\rangle, \alpha_R,\alpha_I)$, where $\langle \dots\rangle$ denotes the expectation values, and $\alpha_{R(I)}$ is the real (imaginary) part of $\langle \hat a \rangle$.
In the superradiant phase and for mean-field energies $E>E_{\mathrm{c}}$, two trajectories initially separated by $\Delta \mathbf{\vec x}(0)$ in phase-space diverge as $\vert \Delta \mathbf{\vec x}(t) \vert \sim \vert \Delta \mathbf{\vec x}(0) \vert e^{\lambda_{\mathrm{L}} t}$ at sufficiently long times \cite{strogatzbook}.  The exponential growth, associated with  a positive Lyapunov exponent $\lambda_{\mathrm{L}} >0$, diagnoses chaos in a classical system.
In Fig.~\ref{fig:phase}b we show the maximal Lyapunov exponent for an ensemble of random initial product states
as a function of the  transverse field and the normalized mean-field energy $E/E_{\mathrm{c}}$ (see Methods). For $E<E_{\mathrm{c}}$ in the superradiant phase ($B<B_{\mathrm{c}}$) and all energies in the normal phase  ($B>B_{\mathrm{c}}$), the  Lyapunov exponent is  small or zero, consistent with the Poissonian  character  of the quantum level statistics in this parameter regime\cite{Emary2003_PRL,ChavezCarlos2016}. For $E>E_{\mathrm{c}}$ and $B<B_{\mathrm{c}}$  a positive exponent is found signaling chaos. Note that the state  $\vert\Psi_0^c\rangle=\vert (-N/2)_x\rangle \otimes \vert 0\rangle$, where $\hat S_x \vert (-N/2)_x\rangle=\left(-N/2\right) \vert (-N/2)_x\rangle$, lies exactly at the ESQPT critical energy,  $\langle \Psi_0^c \vert \hat H_D\vert \Psi_0^c \rangle =E_{\mathrm{c}}$, and possesses the largest classical $\lambda_{\mathrm{L}}$ (see Fig.~\ref{fig:phase}).

%{\it \bf  Quantum  chaos}
In quantum systems OTOCs may serve as a diagnostic for quantum chaos. However, such diagnosis has proved difficult, since any exact numerical treatment is only possible in small systems, where many-body observables saturate quickly at the  \emph{Ehrenfest} time given by $\lambda_{\mathrm{Q}} t^*\sim \log N$, at which  the quantum information is  thoroughly lost to a ``local'' observer. \Rev{Here we demonstrate that we can overcome this limitation and compute OTOCs for macroscopic systems if,
for a   Hermitian  operator $\hat{G}$,
one restricts  $\hat W_{G}=e^{i\delta \phi \hat{G}}$ to be a  sufficiently small perturbation ($\delta \phi \ll 1$)  and sets $\hat V$ to be a projection operator onto a simple initial  state $\vert \Psi_0 \rangle$, i.e.  $\hat V= \hat \rho(0)=\vert \Psi_0 \rangle \langle \Psi_0\vert $. This is  because in the perturbative limit $\delta \phi \ll 1$, this particular type of fidelity OTOC (FOTOC) \cite{Garttner2016,GarttnerMQC2017} ,  $\mathcal{F}_G(t,\delta \phi)\equiv \langle \hat{W}_{G}^\dagger(t) \hat{\rho}(0) \hat{W}_{G}(t) \hat{\rho}(0)\rangle$ (such that for a pure state $\mathcal{F}_G(t) \equiv \vert \langle \psi_0 \vert e^{i\hat{H}t} e^{i\delta\phi\hat{G}} e^{-i\hat{H}t} \vert \psi_0 \rangle \vert^2$) reduces to \cite{Schmitt2018} }
\begin{align}
 1-\mathcal{F}_G(t,\delta \phi) & \approx
 \delta \phi^2 \Big (\langle \hat G^2(t)\rangle -  \langle \hat G(t)\rangle^2\Big)\equiv\delta \phi^2 \mathrm{var}[\hat{G}(t)] ,
\end{align}
where $\mathrm{var}[\hat{G}(t)]$ is the variance of $\hat{G}$. This relation establishes a connection between the exponential growth of quantum variances and quantum chaos, enables  us to visualize the scrambling dynamics of a quantum system using a semi-classical picture \cite{Fox1994,PhysRevLett.121.210601,Fine_2014,Fine_2015} and to map the FOTOC to a two-point correlator which
%what about a Fock state -- that falls into uncorrelated but is hard to sample formally:%, for initial uncorrelated states,
can be computed using well known phase-space methods, such as the truncated Wigner approximation (see Methods)~\cite{polkovnikov2010,Johannes2015}.
We observe perfect agreement between the exact dynamics of the FOTOC with the associated variance, $ \mathrm{var}(\hat{G})$ for sufficient small $\delta \phi$, enabling us to use phase-space methods to compute the FOTOCs in a parameter regime inaccessible  to exact numerical diagonalization where exponential scrambling can be clearly identified.

Moreover, it provides a link between the FOTOCs and the quantum Fisher information (QFI) \cite{GarttnerMQC2017,Toth2014, Pezze2016, Macri_Loschmidt_2016}, as the variance of $\hat G$ is proportional to the QFI of a pure state, whilst for a mixed state the variance gives a lower bound on the QFI. \Rev{Note that in the latter case FOTOCs are defined by replacing $\hat{V}$ by the initial density matrix $\vert \Psi_0 \rangle \langle \Psi_0 \vert \to \hat{\rho}_0$, and expectation values are computed by appropriate traces.} The QFI quantifies the maximal precision with which a parameter $\delta\phi $ in the unitary of $\hat W$ can be estimated using an interferometric protocol with an input quantum state $\ket{\Psi(t)}$, while simultaneously serving as a witness to multipartite entanglement  \cite{Pezze2009,Hyllus2012,Toth2012,GarttnerMQC2017}.

In Fig.~\ref{fig:ChaosOTOC}a we plot the FOTOCS of a small perturbation using $\hat G=\hat X=\frac12(\hat a + \hat a^\dagger)$  starting with   $\vert\Psi_0\rangle=\vert\Psi_0^c\rangle$.  In the superradiant phase we observe that after a short time of slow dynamics, $t_\lambda\sim \lambda_{\mathrm{Q}}^{-1} $, the FOTOCs feature an exponential growth $ \sim e^{\lambda_{\mathrm{Q}} t}$, before saturating at $t^* \sim \log N$ (see inset). The quantum exponent is found to be independent of system size $N$.
For this initial state, and all the product states we have investigated numerically (Supplementary Note 1), we have observed that $\lambda_{\mathrm{Q}} \simeq 2\lambda_{\mathrm{L}}$, as shown in Fig.~\ref{fig:ChaosOTOC}b. Indeed, for any $\hat{G}$ which corresponds to a linear function of the classical phase-space variables (see Methods and Supplementary Note 1) the quantum exponent should be related to the classical Lyapunov exponent by this relation. A similar factor of two relating the classical and quantum exponents has previously been observed in Refs.~\cite{Schmitt2018, Galitski}. This correspondence can be explained by semi-classical arguments (see Methods), and the numeric prefactor is attributable to the definition of the classical Lyapunov exponent in terms of a distance in phase-space, whilst the FOTOC reduces to the quantum variance.

%However, we expect deviations to occur when the initial state is not classically meaningful, e.g., not a spin or bosonic coherent state. In this vein, we also investigated the growth of OTOCs for thermal ensembles for which a quantum bound on the observable exponent has been conjectured \cite{}. while our numerical calculations have failed to demonstrate exponential growth from such thermal ensembles, this could be attributed to the small system sizes and low-temperatures to which we are computationally restrained. In particular, the latter implies a large projection onto the low energy subspace of the Dicke model which is known to remain quasi-integrable in the superradiant phase. A pertinent question to explore in the future is thus whether more general bounds could be placed on the quantum exponent, without restrictions on the nature of the initial state.

%{ \it \bf Connections between OTOCs and entanglement}

\subsection{FOTOCs as a probe of entanglement and quantum thermalization}
We now move beyond the semi-classical arena and explore connections between FOTOCs and entanglement entropy.
In a closed system $\rm{S}$ the second-order RE
$S_2(\hat\rho_A)=-\log  \mathrm{Tr}(\hat{\rho}_A^2)$  measures the entanglement between a subsystem $A$ and its complement $A_c= \rm{S}-A$, with $\hat{\rho}_A$ the reduced density matrix of $A$ after tracing over $A_c$. Although scrambling and entanglement buildup are closely connected they are not the same. Nevertheless, a formal relationship between the OTOCs and $S_2(\hat\rho_A)$ exists \cite{Hosur2016}, which requires averaging OTOCs correlators over a complete basis of operators of the system subsystem $A$. Based on this relation,
measuring RE via OTOCs appears as challenging as directly measuring  $S_2(\hat\rho_A)$.
However, this is  not always the case. We will show that for collective Hamiltonians, such as the DM, there is a simple correspondence between the Fourier spectrum of FOTOCs and the RE, which facilitates experimental access to $S_2(\hat\rho_A)$ via global measurements and collective rotations.

To illustrate the connection we first write the density matrix of the full system   in a basis spanned by the eigenstates of the spin operator $\hat{S}_{\bf r}\equiv ({\bf e}_{\bf r}\cdot \hat{ \vec {\bf S}})$ where  ${\bf e}_{\bf r}$ is a unit vector in the Bloch sphere, satisfying
$\hat{S}_{\bf r}\vert m_{\bf r} \rangle = m \vert m_{\bf r} \rangle$, and
$\hat{n} \vert n\rangle=  \vert n\rangle$ the mode number operator $\hat{n}=\hat{a}^\dagger \hat a$, i.e.  $\hat{\rho}= \sum_{\substack{n,n^{\prime} \\ m_{\bf r}, m_{\bf r}^{\prime} }} \varrho^{n^{\prime},n}_{m_{\bf r}^{\prime},m_{\bf r}} \vert n^{\prime}\rangle \langle n \vert \otimes \vert m_{\bf r}^{\prime} \rangle \langle m_{\bf r} \vert $.  We adopt a convention for the coefficients of the density matrix elements where superscripts are associated with the bosonic mode, and subscripts with the spin. In this basis the density  matrix can be  divided into  blocks,  $\hat{\rho}=\sum_M \hat{\rho}^{\hat{S}_{\bf r}}_M$ with $\hat{\rho}^{\hat{S}_{\bf r}}_M= \sum_{\substack{n,n^{\prime} \\ m_{\bf r} }} \varrho^{n^{\prime},n}_{m_{\bf r}+M,m_{\bf r}} \vert n^{\prime}\rangle \langle n \vert \otimes \vert m_{\bf r}+ M \rangle \langle m_{\bf r} \vert $, in  such a way that $\hat{\rho}^{\hat{S}_{\bf r}}_M $ contains all coherences between states with spin eigenvalues that differ by $M$. A similar decomposition can be performed in terms of the bosonic coherences as $\hat{\rho}=\sum_M \hat{\rho}^{{\hat n}}_M$ with $\hat{\rho}^{\hat {n}}_M= \sum_{\substack{n \\ m_{\bf r},m^{\prime}_{\bf r}}} \varrho^{n+M,n}_{m^{\prime}_{\bf r},m_{\bf r}} \vert n+M\rangle \langle n \vert \otimes \vert m^{\prime}_{\bf r} \rangle \langle m_{\bf r} \vert$. Associated with this representation one can define the  so called  multiple quantum intensities $I^{\hat {G}}_M= \mathrm{Tr} [ \hat{\rho}^{\hat G}_{-M} \hat{\rho}^{\hat G}_M]$. Of particular interest for us are the $I^{\hat {G}}_0$ components which are ``incoherent'' with respect to $\hat G$.

The intensities $I^{\hat {G}}_M(t)$ can be accessed experimentally from FOTOCs via the relation ${\mathcal F}_G(t,\phi)=\sum_M I^{\hat {G}}_M(t) e^{-i M \phi}$\cite{Baum1985, Alvarez2015, Sanchez2009} \cite{Garttner2016,GarttnerMQC2017} by choosing $\hat{W}_G(\phi)= e^{-i \phi  \hat G}$ and $\hat G={\hat S}_{\bf r}$ or $\hat G=\hat n$, i.e. collective spin or boson rotations respectively. In terms of the $I^{\hat {G}}_M(t)$ the entanglement between the spins and the phonons characterized by the purity $\mathrm{Tr}\left[ \hat{\rho}_{ph}^2 \right]=\sum_{\substack{n,n^{\prime} \\ m_{\bf r},m^{\prime}_{\bf r}}} \varrho^{n,n^{\prime}}_{m_{\bf r},m_{\bf r}} \varrho^{n^{\prime},n}_{m^{\prime}_{\bf r},m^{\prime}_{\bf r}}$ can be written as
\begin{equation}
 \mathrm{Tr}\left[ \hat{\rho}_{\mathrm{ph}}^2 (t)\right] \equiv I^{{\hat S}_{\bf r}}_0 (t)+ I^{\hat{n}}_0(t) -  D^{{\hat S}_{\bf r}, \hat{n}}_{{\mathrm{diag}}}(t) + {C}^{{\hat S}_{\bf r}, \hat{n}}_{ {\mathrm{off}}}(t). \label{eqn:EE_MQC}
\end{equation}
The terms $D^{\hat{n},{\hat S}_{\bf r}}_{{\mathrm{diag}}}(t)$ and ${ C}^{\hat{n},{\hat S}_{\bf r}}_{ {\mathrm{off}}}(t)$ are explicitly detailed in the Methods, but importantly $D^{\hat{n},{\hat S}_{\bf r}}_{{\mathrm{diag}}}(t)$ is composed purely of the diagonal elements of $\hat{\rho}$ while ${ C}^{\hat{n},{\hat S}_{\bf r}}_{ {\mathrm{off}}}(t)$ contains information about coherences. During unitary evolution the characteristic dephasing time of the coherences is $t_c\sim \lambda_{\mathrm{Q}}^{-1}$, which for scrambling systems is much faster than $t^*\sim \lambda_{\mathrm{Q}}^{-1}\log{N}$. After $t_c$ any remaining coherences are fully randomized and destructively interfere yielding ${C}^{{\hat S}_{\bf r}, \hat{n}}_{ {\mathrm{off}}}\to0$. This feature, together with the fact that for those systems also the magnitude of $D^{{\hat S}_{\bf r}, \hat{n}}_{{\mathrm{diag}}}$ becomes much smaller than $I^{{\hat S}_{\bf r}}_0 $ and $I^{\hat{n}}_0$ as the density matrix spreads out over the systems degrees of freedom, allows us to approximate $\mathrm{Tr}\left[ \hat{\rho}_{\mathrm{ph}}^2 (t)\right] \approx I^{{\hat S}_{\bf r}}_0 (t)+ I^{\hat{n}}_0(t)$. While at $t<t_c$ these conditions are not necessarily satisfied, we still find that there can be a correspondence between the FOTOCs and RE by picking a state that is fully incoherent at time $t=0$, ${C}^{{\hat S}_{\bf r}, \hat{n}}_{ {\mathrm{off}}}(0)=0$.  An example of such a state is $|\Psi_0^c\rangle$ and  $\hat{G}=\hat{S}_x$. This choice enforces the ${C}^{{\hat S}_{\bf r}, \hat{n}}_{ {\mathrm{off}}}$ term to remain small at short times. Moreover, for $|\Psi_0\rangle$ we find it is also possible to access $S_2(\hat{\rho}_{\mathrm{ph}})$ via ${I}^{\hat S_{\bf r}}_0$ even in the regime $B>B_{\mathrm{c}}$, where no scrambling occurs. This is because the contributions from $I^{\hat n}_0$ and $D^{{\hat S}_{\bf r}, \hat{n}}_{{\mathrm{diag}}}$ cancel and $\mathrm{Tr}\left[ \hat{\rho}_{\mathrm{ph}}^2 (t)\right] \approx I^{\hat S_{\bf r}}_0$.

In Fig.~\ref{fig:ThermErg}a  we show the typical behaviour of the RE, $S_2(\hat \rho_{\rm ph})$, in the two different phases for  $|\Psi_0\rangle$.
First, in the normal phase [panel (i)], $B \gg B_{\mathrm{c}}$, the dynamics is dominated by precession about the transverse field and the entanglement entropy exhibits small amplitude oscillations~\cite{Wall2017}. Conversely, in the superradiant phase [panel (ii)] $B \ll B_{\mathrm{c}}$ we observe a rapid growth of entanglement and saturation past the transient regime. We summarize our results in Fig.~\ref{fig:ThermErg}b where we plot the time-averaged value of $S_2\left(\hat \rho_{\rm ph}\right)$ vs $\sqrt{B_{\mathrm{c}}/B}$. We associate the fast growth of $S_2\left(\hat \rho_{\rm ph}\right)$ at $B/B_{\mathrm{c}}\sim1$ with a cross-over from the integrable to the chaotic regime. To further illustrate this connection, we compare the approximate RE obtained via $S^{\hat{S}_{\mathbf{r}}, \hat{n}}_{\mathrm{F}}\equiv-\mathrm{log}[I^{{\hat S}_{\bf r}}_0 (t)+ I^{\hat{n}}_0(t)]$ and $S^{\hat{S}_{\mathbf{r}}}_{\mathrm{F}}\equiv-\mathrm{log}[I^{\hat S_{\bf r}}_0]$ with the exact RE in Fig.~\ref{fig:ThermErg}b.
%It is observed that in all parameter regimes one can make a quantitative link between the RE and FOTOCs especially under proper optimization of the rotation direction ${\bf e}_{\bf r}$ at each time (see Methods).
It is observed that in all parameter regimes one can make a quantitative link between the RE and FOTOCs, especially under proper optimization of the rotation axis $\hat{S}_{\mathbf{r}}$ at each time to minimize the coherence and diagonal terms in Eq.~(\ref{eqn:EE_MQC}) (see Methods and Supplementary Methods).
%We observe excellent agreement in all regimes.%, especially for $B \ll B_c$ when $\mathrm{Tr}\left[ (\hat{\varrho}_{\mathrm{ph}}^{II}(t))^2 \right]$  holds best.

The saturation of $S_2(\hat{\rho}_{\mathrm{ph}})$ for $B<B_{\mathrm{c}}$ is a signature  of thermalization. One can test how ``thermalized'' the quantum system is by comparing the behavior of the spin and phonon distributions in the long time limit %steady state
with those of the corresponding diagonal ensemble, characterized by a mixed density matrix $\hat \rho_D$ with purely diagonal elements (see Methods) \cite{DAlessio2016,Nandkishore2015,Eisert2016}. These comparisons are shown in Fig.~\ref{fig:ThermErg}c, where the time evolved distributions and the ones drawn from the  diagonal ensemble are almost indistinguishable.

We can also investigate the growth of entanglement on different size bi-partitions for $B<B_{\mathrm{c}}$. For that we split the spin system into a subsystem  of size $L_A\leq N$ and evaluate  $S_2(\hat \rho_{L_A})$  by computing the reduced density matrix $\hat \rho_{L_A}$ by  tracing over the bosonic degree of freedom and the remaining $N-L_A$ spins. To demonstrate the entanglement grows with system size in a manner consistent with an equivalent thermal state we plot the predictions of a canonical ensemble (see Methods). We observe volume-law entanglement growth for $L_A\ll N$  (see Fig.~\ref{fig:ThermErg}d ). However, for $L_A\sim N$ the entanglement growth deviates from this simple prediction. These deviations occur as the full state of the system is pure, and thus eventually one needs to recover $S_2(\hat{\rho}) = 0$, requiring a negative curvature. To  demonstrate the intertwined nature of thermalization and the build-up of entanglement  we plot the predictions of a canonical ensemble  indicated by the dotted purple line (see Methods). We note that FOTOCs can also be used to probe this scaling of the RE with subsystem size. To this end, both $\hat{V}$ and $\hat{W}$ should be restricted to a partition of size $L_A$ of the system, but otherwise the corresponding multiple quantum intensities are computed as discussed above (see also Methods). Figure~\ref{fig:ThermErg}d shows the excellent agreement between the partial system FOTOCs (blue squares) and RE (black diamonds), comparisons that  illustrate the utility of FOTOCs to characterize complex many-body entanglement.

\subsection{Experimental implementation in trapped ion simulators}

\begin{figure}
 \includegraphics[width=12cm]{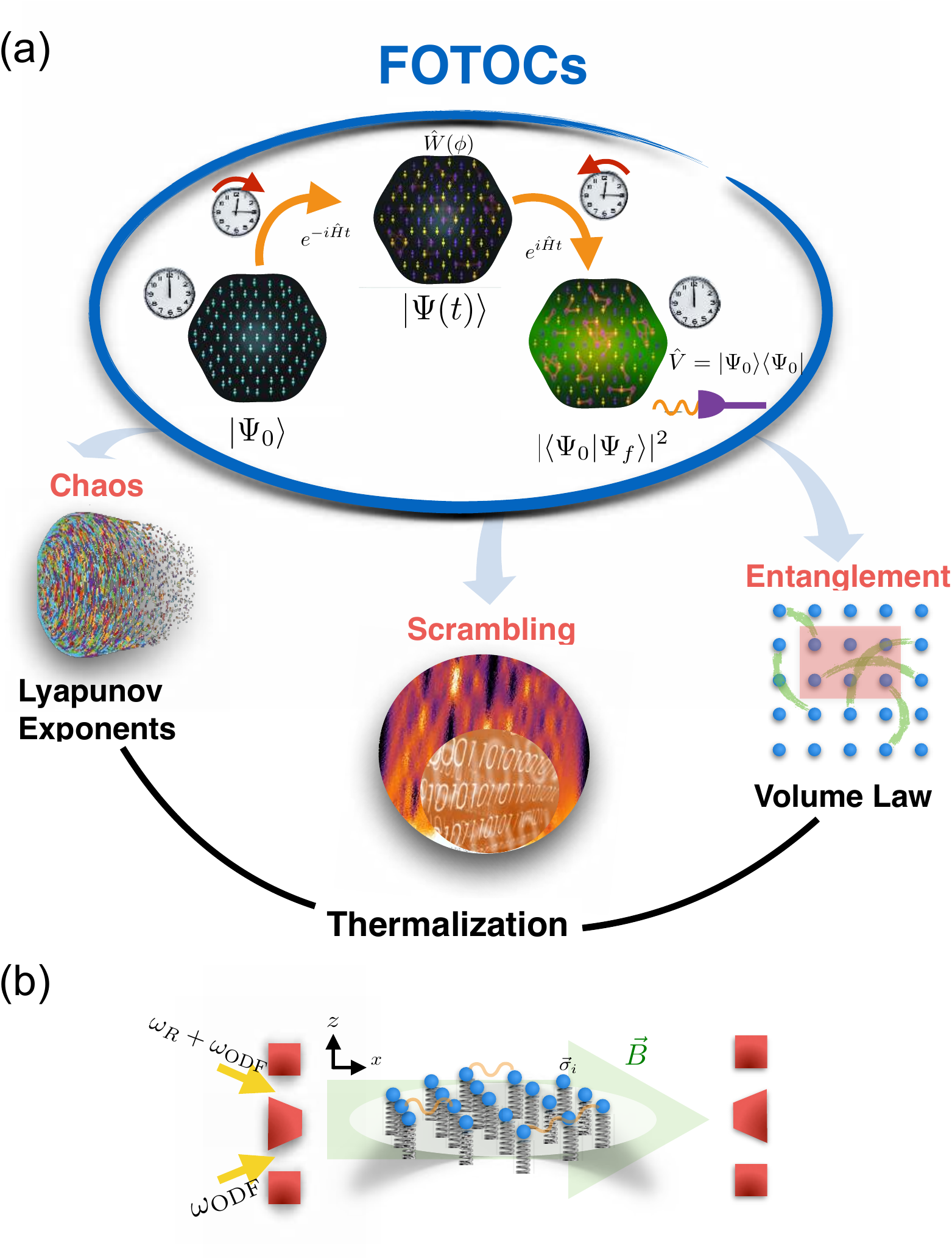}
 \caption{   Unifying chaos, scrambling, entanglement and thermalization through the measurement of fidelity out-of-time-order correlators (FOTOCs).  Scheme: an initial  state, $\ket{\Psi_0}$ is evolved under an interacting Hamiltonian $\hat H$ for a time $t$. Inverting the sign of $\hat H$  and evolving again for time $t$ to the final state $\ket{\Psi_f}$, implements the many-body time-reversal, which ideally takes the system back to the initial state  $\ket{\Psi_0}$. If a perturbation $\hat{W}(\phi)$ is inserted between the two halves of the time evolution and the many-body overlap with the initial state is measured at the end of the protocol, $\hat{V}=\ket{\Psi_0} \bra{\Psi_0}$, then a special type of \emph{fidelity} OTOC (FOTOC) is implemented.
 (b) The Dicke model is engineered in a Penning trap ion crystal by applying a pair of lasers, resonant only with the center of mass mode, to generate the spin-phonon interaction and resonant microwaves to generate the transverse field.}
 \label{fig:SchemFig}
\end{figure}

\begin{figure}
 \includegraphics[width=16cm]{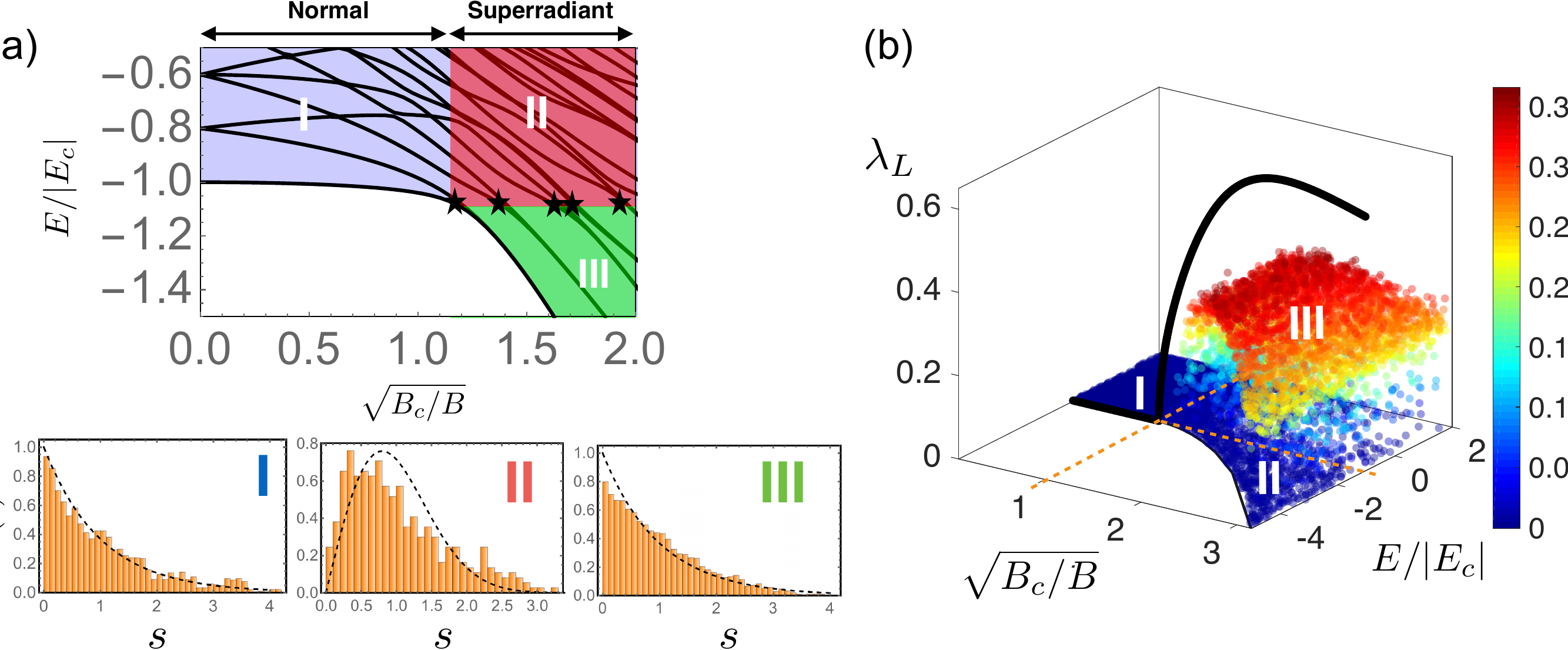}
 \caption{Characterization of classical and quantum chaos in the Dicke model. (a) Phase diagram of the Dicke model. At zero temperature it exbibits a quantum phase transition between a normal to a superradiant phase, at $B=B_{\mathrm{c}}$. A line of excited energy quantum phase-transitions (ESQPTs) occurs at the critical energy $E_{\mathrm{c}} = -BN/2$, signaled by singularities in the energy level structure (indicated by stars). Note that for figure clarity we have used a small system $N = 20$  resulting in the small deviation of the ESQPTs from $E_{\mathrm{c}} = -BN/2$. The ESQPTs are accompanied by a change in the level statistics which we denote by (I)-(III) (note that no eigenstates exist in the unlabelled white region). For (II) and (III) the spectrum is divided into low and high energy parts, separated by the ESQPT at $E=E_{\mathrm{c}}$, from which the statistics $P(s)$, where $s$ is the level spacing, are computed separately. (I) exhibits Poissonian statistics (regular regime), while (II) displays statistics similar to a Wigner-Dyson distribution indicative of level repulsion and quantum chaos, and (III) exhibits a mixture of both. The numerical parameters are $g/(2\pi)=0.66$~kHz and $\delta/(2\pi)=0.5$~kHz. (b) Lyapunov exponents for the mean-field dynamics of an ensemble of random states sorted by normalized mean-field energy $E/\vert E_{\mathrm{c}}\vert$ with $E_{\mathrm{c}} = -BN/2$, as a function of the field $\sqrt{B_{\mathrm{c}}/B}$ relative to critical field $B_{\mathrm{c}}=4g^2/\delta$. A crossover between regular ($B > B_{\mathrm{c}}$) and chaotic dynamics ($B < B_{\mathrm{c}}$) characterized by $\lambda \simeq 0$ and $\lambda_{\mathrm{L}} > 0$ respectively, occurs at $B=B_{\mathrm{c}}$. For $B>B_{\mathrm{c}}$ and energies $E \lesssim E_{\mathrm{c}}$ the dynamics becomes increasingly regular. Source data are provided as a Source Data file.}
 \label{fig:phase}
\end{figure}

Trapped ions present a promising experimental platform for the investigation of the physics discussed here~\cite{Blatt2017, Monroe53Q, Dickeionpaper2018}.
Here we focus on two-dimensional arrays in a Penning trap where a tunable coupling between the ion's spin, encoded in two hyperfine states, and the phononic center-of-mass (COM) mode of the crystal can be  implemented by a pair of lasers with a beatnote frequency detuned by $\delta $ from   the  COM mode and far from resonance to all other modes, which remain unexcited (Fig.~\ref{fig:SchemFig}b). In the presence of microwaves (which generate the transverse field) resonant with the spin level splitting, the effective Hamiltonian is of the form of Eq.~\ref{eq:Hdicke} as benchmarked in Refs. \cite{Dickeionpaper2018,Cohn2018}. The dynamical control of the transverse field and sign of the detuning from the COM mode enables straightforward implementation of a time-reversal protocol to measure FOTOCs \cite{GarttnerMQC2017} (see Fig.~\ref{fig:SchemFig}). Additionally, the many-body echo  requires the application of a spin echo $\pi$ pulse along ${\bf e}_{\bf r}=\hat{y}$ which reverses the signs of $\hat{S}_x$ and $\hat{S}_z$  simultaneously.

Our proposal requires the ability for measuring the fidelity of the full spin-phonon state, which we have not yet demonstrated experimentally. However, this will be possible through a generalization of the protocol discussed in reference~\cite{SchmidtFid} (see Methods). Additionally, our proposal can be adversely affected by decoherence present in the experiment. However, the impact of decoherence will be minimized in future experiments by increasing the magnitude of relevant couplings of the DM via parametric amplification of the ions' motion\cite{Dickeionpaper2018,FossFeigPA}, thus reducing the ratio of dissipative to coherent evolution. We illustrate the predicted effect of decoherence, which is dominated by single-particle dephasing due to light scattering from the lasers, in the inset of Fig.~\ref{fig:ThermErg}b. We include the enhancement of the coherent parameters via the protocol described in~\cite{FossFeigPA} while using the typical experimental decoherence rate of $\Gamma=60$~s$^{-1}$. The single-particle decoherence is modelled by an exponential decay of the FOTOC components $I^{\hat{G}}_0 \to I^{\hat{G}}_0 e^{-\Gamma N t}$ (see Methods). The numerical calculation indicates that even with decoherence the crossover between the two regimes at $B\sim B_{\mathrm{c}}$ is still well captured. Due to numerical complexity of solving a master equation we restrict our simulations to $N=40$ ions.

\section{Discussion}

We have demonstrated that FOTOCs connect the fundamental concepts of scrambling, chaos, quantum thermalization, and multipartite entanglement in the DM. While the concepts presented here have been limited to collective Hamiltonians we believe they can be generalized to more complex many-body models \Rev{(Supplementary Note 2)}. For example, FOTOCs could provide an alternative approach for performing efficient measurements of RE in a way comparable to other state-of-the-art methods which have been used to probe entanglement in systems with up to twenty ions \cite{Roos2018}. Generically, FOTOCS could serve as a new experimental tool capable of uncovering bounds on information transport and computational complexity,  and shed light on how  classical behaviors in macroscopic systems  emerge from purely microscopic quantum effects.

%Last line needs to be better. Want to convey the idea of FOTOCs as a tool for concepts which are now connecting large swathes of physics.

\newpage
{\it \bf Methods}

\begin{figure}
 \includegraphics[width=16cm]{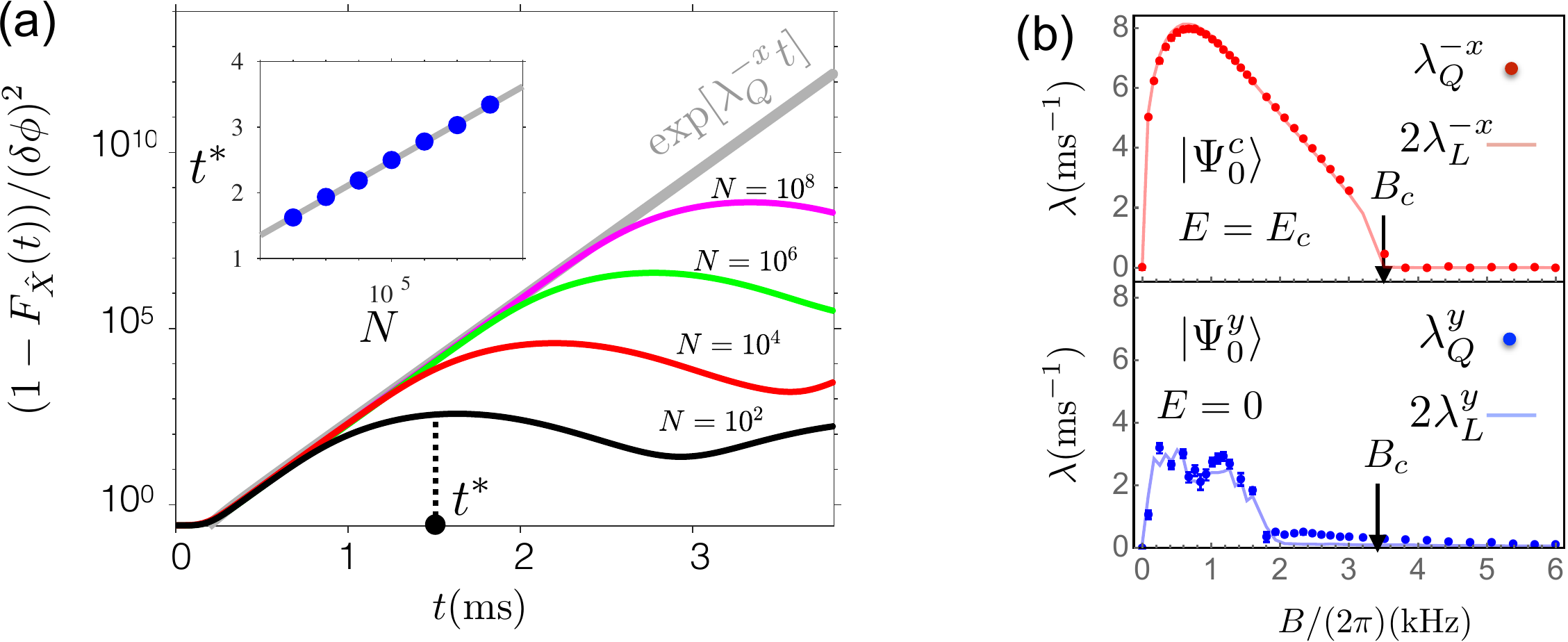}
 \caption{Signatures of classical chaos in quantum FOTOCs. (a) Initial exponential growth of the FOTOC, $[1-\mathcal{F}_X(t)]/(\delta\phi)^2$ and the initial state $\vert\Psi_0^c\rangle=\vert (-N/2)_x\rangle \otimes \vert 0\rangle$ (see Supplementary Note 1 for examples of exponential growth in other states). \Rev{We assume $\delta\phi \ll 1/N$ such that we may equivalently use $\mathrm{var}(\hat{X}) \simeq [1-\mathcal{F}_X(t)]/(\delta\phi)^2$ for the plotted data.} The scrambling time $t^*$ is defined by the saturation of the FOTOC, which we extract from the first maximum and plot in the inset (blue data). We find $t_{*} \sim a_0 + \mathrm{log}(N)/\lambda_{\mathrm{Q}}$ with $a_0$ a fit parameter (grey line).
 (b) Lyapunov exponent, $\lambda$, as a function of transverse field: Quantum  $\lambda_{\mathrm{Q}}$ (red markers) and classical $2\lambda_{\mathrm{L}}$ (solid lines). Superscript notation of the exponents denotes the initial polarization of the chosen coherent spin state. Top panel for $\vert\Psi_0^c\rangle$, the same state as (a), and bottom for  $ \vert \Psi^y_0 \rangle\vert (-N/2)_y\rangle \otimes \vert 0\rangle$ , here  $N=10^4$ particles. In both plots we observe $\lambda_{\mathrm{Q}} \simeq 2\lambda_{\mathrm{L}}$. Error bars for $\lambda_{\mathrm{Q}}$ are a $95\%$ confidence interval from an exponential fitted to the numerical data. Coupling $g$ and detuning $\delta$ are same as Fig.~\ref{fig:phase}. In (a) $B/(2\pi)=0.7$~kHz ($B/B_{\mathrm{c}} = 0.2$). Source data are provided as a Source Data file.}
 \label{fig:ChaosOTOC}
\end{figure}

\noindent{\it \bf Classical dynamics and equations of motion}
The results presented for the classical model in Fig.~\ref{fig:ChaosOTOC} are obtained from the Heisenberg equations of motion for the operators via a mean-field ansatz, wherein the operators are replaced by the c-number expectation values, i.e., $\hat{S}_{j} \to \langle \hat{S}_{j} \rangle$ for $j = x,y,z$ and $\hat{a} \to \langle \hat{a} \rangle$ [where we adopt $\alpha_R(I)$ as the real (imaginary) component of $\langle \hat{a} \rangle$]. 
% Our analysis is further simplified by the introduction of the re-normalized variables
% \begin{equation}
%     \alpha_{R} = \frac{\mathrm{Re}(\langle \hat{a} \rangle)}{\sqrt{N}} , \quad \quad 
%     \alpha_{I} = \frac{\mathrm{Im}(\langle \hat{a} \rangle)}{\sqrt{N}} ,
% \end{equation}
% and
% \begin{equation}
%     l_{j} = \frac{2\langle \hat{S}_{j}\rangle}{N} . 
% \end{equation}
% We thus obtain an equation of motion for $\vec{x} = (\alpha_{R}, \alpha_{I}, l_x, l_y, l_z)$, 
% \begin{equation}
%     \frac{d\vec{x}}{dt} = F(\vec{x}) , \label{eqn:dx_dt}
% \end{equation}
% where
% \begin{align}
%     F(\vec{x}) &= \begin{pmatrix}
%           -\delta \alpha_I \\
%           \delta \alpha_R + gl_z \\
%           4g\alpha_R l_y \\
%           -Bl_z - 4g\alpha_Rl_x \\
%             B l_y
%          \end{pmatrix} .
% \end{align}
We thus obtain an equation of motion for $\mathbf{\vec x}=(\langle \hat S_x\rangle , \langle \hat S_y \rangle , \langle \hat S_z\rangle, \alpha_R,\alpha_I)$, 
\begin{equation}
    \frac{d\mathbf{\vec{x}}}{dt} = F(\mathbf{\vec{x}}) , \label{eqn:dx_dt}
\end{equation}
where
\begin{align}
    F(\mathbf{\vec{x}}) &= \begin{pmatrix}
           -\delta \alpha_I \\
           \delta \alpha_R - \frac{2g}{\sqrt{N}}\langle \hat S_z \rangle \\
           -\frac{4g}{\sqrt{N}}\alpha_R \langle \hat{S}_y \rangle \\ 
           -B \langle \hat{S}_z + \frac{4g}{\sqrt{N}}\alpha_R\langle \hat{S}_x \rangle \\
            B \langle \hat{S}_y \rangle
         \end{pmatrix} .
\end{align}

\noindent{\it \bf Lyapunov exponent}
The existence of classical chaos can be characterized by the Lyapunov exponent $\lambda_{\mathrm{L}}$. By definition, classical chaos implies that two initially close trajectories separated by a distance in phase-space $\Delta\mathbf{\vec{x}}(0) = \vert \mathbf{\vec{x}}_1(0) - \mathbf{\vec{x}}_2(0) \vert$ diverge exponentially, $\vert\Delta\mathbf{\vec{x}}(t)\vert \approx \vert\Delta\mathbf{\vec{x}}(0)\vert e^{\lambda_{\mathrm{L}} t}$, and thus $\lambda_{\mathrm{L}} > 0$ is a signature of chaotic dynamics. 

Formally, the Lyapunov exponent is then defined by taking the limit \cite{strogatzbook}
\begin{equation}
    \lambda_{\mathrm{L}} \equiv \lim_{t\to \infty} \lim_{\vert\Delta\mathbf{\vec{x}}(0)\vert \to 0} \frac{1}{t} \mathrm{log} \frac{\vert\Delta\mathbf{\vec{x}}(t)\vert}{\vert\Delta\mathbf{\vec{x}}(0)\vert} . \label{eqn:Lyapunov}
\end{equation}

As the phase-space of our co-ordinate system is bounded we evaluate Eq.~(\ref{eqn:Lyapunov}) using the tangent-space method \cite{ChavezCarlos2016,Skokos2010}. Essentially, rather than monitoring the physical separation $\vert \Delta\mathbf{\vec{x}}(t)\vert$ of a pair of initially nearby trajectories, one can instead solve for the separation in tangent space, denoted by $\delta\mathbf{\vec{x}}(t)$, and substitute this distance into Eq.~(\ref{eqn:Lyapunov}). The tangent-space separation $\delta\mathbf{\vec{x}}(t)$ can be dynamically computed by assuming an infinitesimal initial perturbation to a reference trajectory starting at $\mathbf{\vec{x}}(0) = \mathbf{\vec{x}}_0$, leading to the system of equations
\begin{eqnarray}
    \frac{d\mathbf{\vec{x}}}{dt} & = & F(\mathbf{\vec{x}}) , \label{eqn:dx_dt_sys} \\
    \frac{d\mathbf{\Phi}}{dt} & = & \mathbf{M} \mathbf{\Phi} . \label{eqn:dPhi_dt_sys}
\end{eqnarray}
Here, $\mathbf{\Phi}$ is the fundamental matrix and $M_{ij} \equiv dF_i/dx_j$. The tangent-space separation with respect to the initial point in phase-space $\mathbf{\vec{x}}(0)=\mathbf{\vec{x}}_0$ is extracted by computing $\delta\mathbf{\vec{x}}(t) \equiv \mathbf{\Phi}\delta\mathbf{\vec{x}}(0)$ with $\mathbf{\Phi}(0) = \mathbb{I}$.

As we are only interested in the maximum Lyapunov exponent, it suffices to choose the initial separation $\delta\mathbf{\vec{x}}(0)$ along a random direction in phase-space, and we propagate Eqs.~(\ref{eqn:dx_dt_sys}) and (\ref{eqn:dPhi_dt_sys}) for each  initial condition $\mathbf{\vec{x}}_0$ for sufficiently large $t$ that our estimate of $\lambda_{\mathrm{L}}$ from Eq.~(\ref{eqn:Lyapunov}) converges.

\noindent{\it \bf Connection between classical and quantum Lyapunov exponents}
In our discussion of the exponential growth of FOTOCs, we have argued that $\lambda_{\mathrm{Q}}$ is intimately related to the classical Lyapunov exponent $\lambda_{\mathrm{L}}$. Specifically, we have that $\lambda_{\mathrm{Q}} \simeq 2\lambda_{\mathrm{L}}$. Here, we further articulate this connection using a semi-classical description of the quantum dynamics, specifically by considering the evolution in the truncated Wigner approximation (TWA) \cite{polkovnikov2010}. 

First, we remind the reader that for a small perturbation $\delta\phi$, a FOTOC $\mathcal{F}_G(t,\delta\phi)$ can be expanded to $\mathcal{O}(\delta\phi^2)$ as $\mathcal{F}_G(t,\delta\phi) \approx 1 - \delta\phi^2\mathrm{var}(\hat{G})$. A simple conclusion from this expansion is that if  $\mathcal{F}_G(t,\delta\phi)$ grows exponentially we can attribute this behaviour to the variance, i.e. it must be true that $\mathrm{var}(\hat{G}) \rangle \sim e^{\lambda_{\mathrm{Q}} t}$. 

A semi-classical explanation of this exponential growth is simplified by assuming that $\hat{G}$ is an operator which is linear in the classical phase-space variables $\mathbf{\vec{x}}$. For concreteness, let us consider $\hat{G}=\hat{X}=\frac12(\hat{a}+\hat{a}^{\dagger})$ as in Fig.~\ref{fig:ChaosOTOC} of the main text, which corresponds to $\alpha_R$ in the classical phase-space. 

Next, we consider a description of the the quantum dynamics within the framework of the TWA. Here, the dynamics is computed by solving the classical equations of motion [Eq.~(\ref{eqn:dx_dt})] with random initial conditions sampled from the corresponding Wigner phase-space distribution of the initial state \cite{polkovnikov2010}. Quantum expectation values are then obtained by appropriate averaging over an ensemble of trajectories, e.g., $\langle \hat{X} \rangle \equiv \overline{\alpha_R}$ where the overline denotes a stochastic average. The random sampling of initial conditions serves to model the quantum fluctuations of the initial state. 

For a classically meaningful initial state (i.e. a product of coherent states for the phonon and spin degrees of freedom), the fluctuations in each of the phase-space variables are typically Gaussian and centered around the expectation values of the initial state. A concrete example to illustrate this is the state $\vert\Psi_0^c\rangle=\vert (-N/2)_x\rangle \otimes \vert 0\rangle$ considered in the main text. For each trajectory, the variable $(\alpha_R)_j$ ($j$ denoting the trajectory) for example, is sampled from a Gaussian distribution with mean zero and variance $1/4$. The connection between the quantum dynamics and classical chaos is made by instead considering sampling only the fluctuations $\delta\alpha_R$ about a central classical trajectory, i.e. $(\alpha_R)_j \to \alpha_R^{cl} + (\delta\alpha_R)_j$. 

Solving the dynamics of the central classical trajectory and the ensemble of fluctuations is then identical to the calculation of Eqs.~(\ref{eqn:dx_dt_sys}) and (\ref{eqn:dPhi_dt_sys}), from which the Lyapunov exponent is calculated. In particular, the connection between quantum and classical exponents is finally made clear by evaluating the quantum variance, 
\begin{eqnarray}
    \mathrm{var}(\hat{X})& = & \left( \overline{\alpha_R^2} - \overline{\alpha_R}^2 \right) , \\
    & \equiv & \left( \overline{\delta\alpha_R^2} - \overline{\delta\alpha_R}^2 \right).
\end{eqnarray}
As $\delta\alpha_R$ is evaluated directly from Eq.~(\ref{eqn:dPhi_dt_sys}) then we expect from our previous calculations that $|\delta\alpha_R| \sim e^{\lambda_{\mathrm{L}} t}$ for a generic random perturbation, sampled according to the TWA prescription, in parameter regimes where there is classical chaos. Thus, we extrapolate that the quantum variance will grow like $\overline{\delta\alpha_R^2} - \overline{\delta\alpha_R}^2 \sim e^{2\lambda_{\mathrm{L}} t}$. Inspection of this final result shows that we should expect $\lambda_{\mathrm{Q}} \simeq 2\lambda_{\mathrm{L}}$.

\noindent{\it \bf Connection between FOTOCs and RE} The connection between the FOTOCs and entanglement entropy is best established by first considering the case of the spin-phonon RE $S_2(\hat{\rho}_{\mathrm{ph}})$. We begin by writing the purity  of the reduced density matrix explicity in terms of the elements of the density matrix,
\begin{eqnarray}
 \mathrm{Tr}\left[ \hat{\rho}^2_{\mathrm{ph}}(t) \right] = \sum_{\substack{n,n^{\prime} \\ m_{\bf r},m^{\prime}_{\bf r}}} \varrho^{n,n^{\prime}}_{m_{\bf r},m_{\bf r}}(t) \varrho^{n^{\prime},n}_{m^{\prime}_{\bf r},m^{\prime}_{\bf r}}(t) . \label{eqn:RenyiPh}
\end{eqnarray}

Our insight is that, in the case of a pure global state, the summation in Eq.~(\ref{eqn:RenyiPh}) for the purity of the reduced density matrix can be manipulated and re-expressed as
\begin{equation}
 \mathrm{Tr}\left[ \hat{\rho}^2_{\mathrm{ph}}(t) \right] \equiv I^{\hat{S}_{\mathbf{r}}}_0(t) + I^{\hat{n}}_0(t) - D^{\hat{S}_{\mathbf{r}},\hat{n}}_{\mathrm{diag}}(t) + C^{\hat{S}_{\mathbf{r}},\hat{n}}_{\mathrm{off}}(t)  , \label{eqn:EE_MQC_SOM} 
 \end{equation}
where
\begin{eqnarray}
 D^{\hat{S}_{\mathbf{r}},\hat{n}}_{\mathrm{diag}}(t) = \sum_{\substack{n, \\ m_{\bf r}}} \left[\varrho^{n,n}_{m_{\bf r},m_{\bf r}}(t)\right]^2 , \\
 C^{\hat{S}_{\mathbf{r}},\hat{n}}_{\mathrm{off}}(t) = \sum_{\substack{n\neq n^{\prime} \\ m_{\bf r} \neq m^{\prime}_{\bf r}}} \varrho^{n,n^{\prime}}_{m_{\bf r},m_{\bf r}}(t) \varrho^{n^{\prime},n}_{m^{\prime}_{\bf r},m^{\prime}_{\bf r}}(t) ,
\end{eqnarray}
are the sum of the squared diagonal elements of the density matrix and the sum over the off-diagonal coherences respectively. Thus, we seek to understand when these latter terms can be neglected and thus the purity (and associated entropy) is expressible in terms of only the $I^{\hat{G}}_0$. 

Firstly, there is the case of a large transverse field, $B\gg B_c$ and an initial state which is polarized along the the direction of the transverse field with vacuum occupation, i.e., $\vert\Psi_0\rangle = \vert (\pm N/2)_x\rangle \otimes\vert 0\rangle$. In this case, we expect the collective spin to remain strongly polarized along the field direction. If we choose the FOTOC spin rotation axis to be along that of the initial state and transverse field, $\hat{S}_{\mathbf{r}}= \hat{S}_x$, then we have that $C^{\hat{S}_x,\hat{n}}_{\mathrm{off}}(t) \approx 0$ due to the absence of initial coherences between the spin sectors in this basis, and by similar reasoning $I^{\hat{n}}_0(t) \approx D^{\hat{S}_x,\hat{n}}_{\mathrm{diag}}(t)$. Hence, we expect Eq.~(\ref{eqn:EE_MQC_SOM}) to simplify so that $\mathrm{Tr}[ \hat{\rho}_{\mathrm{ph}}^2(t) ] \approx I^{\hat{S}_x}_0(t)$. Identical reasoning can be applied in the normal phase ($B>B_c$) when the phonon detuning is the largest energy scale, such that $\mathrm{Tr}[ \hat{\rho}_{\mathrm{ph}}^2(t) ] \approx I^{\hat n}_0(t)$. 

The second scenario is closely related to the first. Consider an initial coherent spin state polarized along an arbitrary spin direction and vacuum phonon occupation. For arbitrary transverse field strength and on sufficiently short time-scales $t\lesssim \lambda_{\mathrm{Q}}^{-1}$, then the spin component of the evolved state remains largely polarized along a particular axis dictated by the initial state. Similar to the first scenario, by choosing the spin rotation of the FOTOC, $\hat{S}_{\mathbf{r}}$, to match the polarization of the initial state, then we will have $\mathrm{Tr}[ \hat{\rho}_{\mathrm{ph}}^2(t) ] \approx I^{\hat{S}_{\mathbf{r}}^2}_0(t)$. This is justified as $C^{\hat{S}_{\mathbf{r}},\hat{n}}_{\mathrm{off}}(t)$ and $I^{\hat{n}}_0(t) + D^{\hat{S}_{\mathbf{r}},\hat{n}}_{\mathrm{diag}}(t)$ vanish, as again the state at short times will not have appreciable coherences between different spin sectors in this basis.

Lastly, for a small transverse field, $B\ll B_c$, and beyond short-times $t \gtrsim \lambda_{\mathrm{Q}}^{-1}$ (i.e., beyond the time-scale when the spin state is still strongly polarized and the second scenario is still valid), we expect Eq.~(\ref{eqn:EE_MQC_SOM}) to be well approximated by $\mathrm{Tr}[ \hat{\rho}^2 _{\mathrm{ph}}(t) ] \approx I^{\hat{S}_{\mathbf{r}}}_0(t) + I^{\hat{n}}_0(t)$ for any spin rotation axis $\hat{S}_{\mathbf{r}}$. This is because initially pure states which are sufficiently scrambled after a quench of the system parameters closely resemble so-called canonical pure thermal quantum (cTPQ) states \cite{Nakagawa2018} in a generic basis. 
% This statement is only rigorously true for $t \gtrsim t^*$, although it can be extended less formally to $t \gtrsim \lambda_Q^{-1}$. 
For cTPQ states, the summation over off-diagonal coherences $C^{\hat{S}_{\mathbf{r}},\hat{n}}_{\mathrm{off}}$ vanishes exactly for a sufficiently large system  as the coherences can be considered as random variables \cite{Nakagawa2018}. Moreover, for a typical spin rotation axis $\hat{S}_{\mathbf{r}}$, the cTPQ state will have a spin distribution $P(M_{\hat{S}_{\mathbf{r}}})$ which is largely delocalized implying  that $D^{\hat{S}_{\mathbf{r}},\hat{n}}_{\mathrm{diag}} \sim 1/(N n_{\mathrm{ph}})$ where $n_{\mathrm{ph}}$ is some constant which characterises the spread of the boson number distribution. The term $D^{\hat{S}_{\mathbf{r}},\hat{n}}_{\mathrm{diag}}$ is then typically much smaller in magnitude when compared to the remaining terms $I^{\hat{S}_{\mathbf{r}}}_0$ and $I^{\hat{n}}_0$. This reasoning leads to $\mathrm{Tr}[ \hat{\rho}^2_{\mathrm{ph}}(t) ] \approx I^{\hat{S}_{\mathbf{r}}}_0(t) + I^{\hat n}_0(t)$. Discussion of the sensitivity of these arguments to the rotation direction can be found in Supplementary Methods.

More generally, we can extend these arguments to extract a correspondence with the Renyi entropy of a generic bipartition of the spin-phonon system. Specifically, splitting the system $\mathcal{S}$ into a subsystem $A$: $L$ spin-1/2s, and its complement $A_c$: $N-L$ spin-$1/2$s and the bosonic mode. In the weak-field regime $B \ll B_c$, $\mathrm{Tr}[ \hat{\rho}_A^2 ] \approx I^{A}_0 + I^{A_c}_0$. Here, the terms $I^A_0$ and $I^{A_c}_0$ are obtained as the Fourier amplitudes of fidelity OTOCs for generalized rotations within each subsystem. Specifically, a local rotation $e^{i\phi\hat{S}_{\mathbf{r},A}}$ taken to act on the spin-$1/2$s in the $A$ subsystem, and a joint (but uncorrelated) rotation $e^{i\phi\hat{S}_{\mathbf{r},A_c}} e^{i\theta\hat{a}^{\dagger}\hat{a}}$ of the spins \emph{and} bosons in the complement $A_c$.

\noindent{\it \bf Experimental implementation}
By preparing an initial spin polarized state, recent experiments \cite{Garttner2016} demonstrated it was possible to measure the many-body overlap of the final state with the initial configuration by flourescence detection. The Dicke model, however, includes spin \emph{and} phonon degrees of freedom. 

While the full spin-phonon fidelity measurement has not yet been demonstrated experimentally, such measurement is possible by extending the method in~\cite{SchmidtFid} to a multi-qubit system. In particular, we note that this proposal is comprised of a two step measurement, where we first measure the spin degree of freedom. The probability of all ions being in the dark state (i.e. all in the state $\vert\downarrow \rangle_z$) can be measured with excellent fidelity and has been previously demonstrated~\cite{Garttner2016}. The dark state does not scatter phonons, and as such, this measurement will not change the state of the phonons. Next one can proceed to measure the phonon occupation via the portocol described in~\cite{SchmidtFid}.

Finally, as noted in the main text, we have taken into account the single-particle decoherence present in the experiment. The results presented in the main text accounted for this by approximating the effects of decoherence by an exponential decay, $\bar{I}^{\hat{G}}_0 \to \bar{I}^{\hat{G}}_0 e^{-\Gamma N t}$. We have justified this approximation by comparing to an efficient numerical solution of the full Lindblad master equation \cite{GarttnerMQC2017,Xu2013,PIQs2018} for smaller system sizes ($N=10$). We find that the decoherence is well captured by the approximate model for all transverse field strengths $B$ considered.  

%We must take into account experimentally important sources of single-particle decoherence. The main contribution is single-particle dephasing due to light scattering from the lasers. In the lower panels of Fig.~\ref{fig:ExpFig}b of the main text we show typical examples of the effect of this decoherence on the evolution of the $\bar{I}^{\hat{S}_{\mathbf{r}}}_0$. In these figures we approximate the effects of decoherence by an exponential decay, $\bar{I}^{\hat{G}}_0 \to \bar{I}^{\hat{G}}_0 e^{-\Gamma N t}$. The $N$-fold enhancement to the decay is due to the fact that experimentally the FOTOCs are accessed by a fidelity measurement at the conclusion of the protocol. We have justified this approximation by comparing to an efficient numerical solution of the full Lindblad master equation \cite{GarttnerMQC2017,Xu2013,PIQs2018}. We find that the decoherence is well captured by the approximate model for all transverse field strengths $B$ considered.  

\begin{figure}\centering
 \includegraphics[width=10cm]{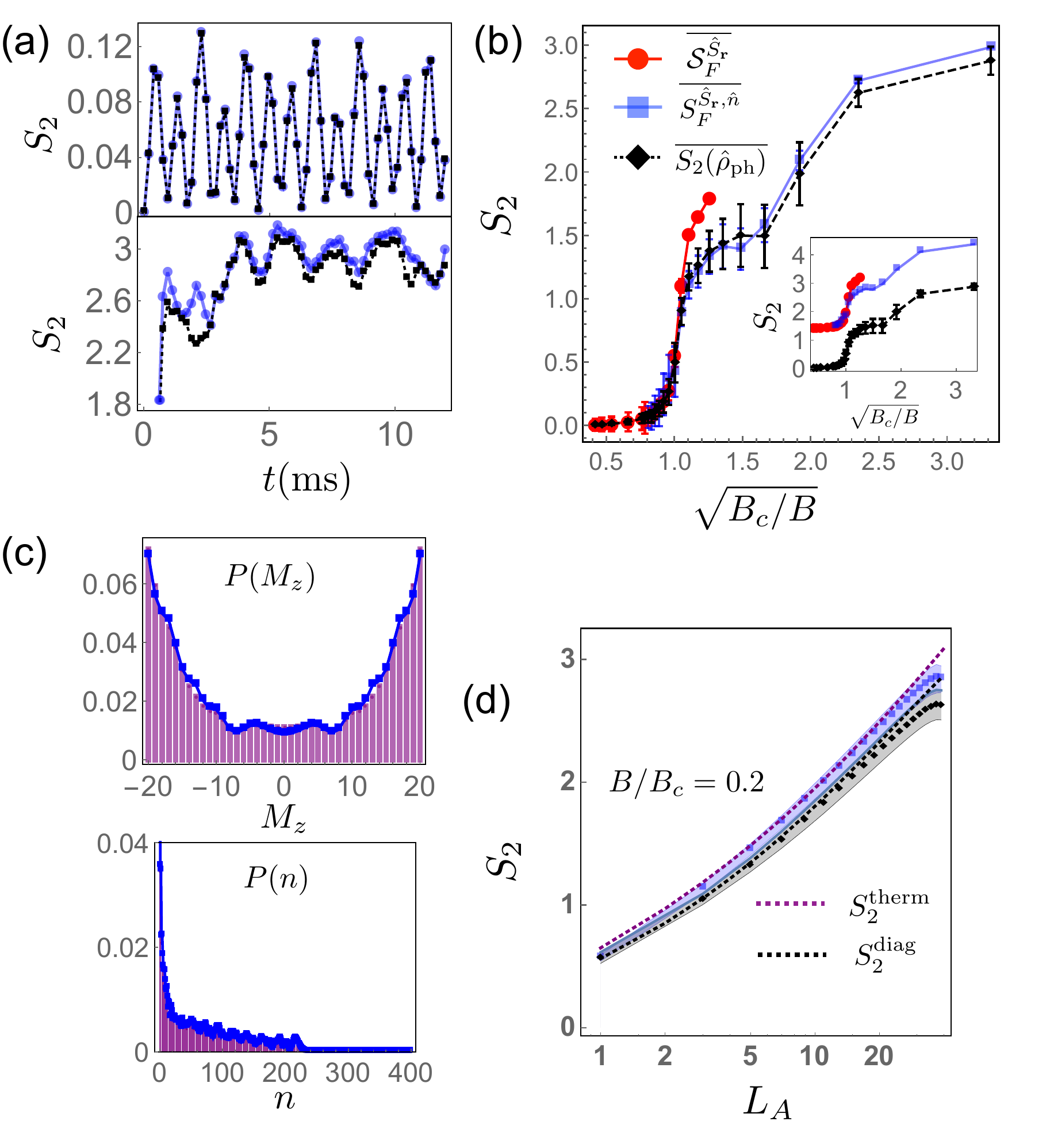}
 \caption{
 Using RE and FOTOCs to characterise chaos and thermalization in the Dicke model. (a) Time evolution of the spin-phonon RE $S_2(\hat{\rho}_{\mathrm{ph}})$ (blue lines) for the initial state $\vert \Psi^c_0 \rangle = \vert (-N/2)_x \rangle\otimes\vert0\rangle$ with $B>B_{\mathrm{c}}$ (top) and $B<B_{\mathrm{c}}$ (bottom). The RE is tracked excellently by the FOTOC expressions $S^{\hat{S}_{\mathbf{r}}}_{\mathrm{F}} = -\mathrm{log}(I^{\hat{S}_{\mathbf{r}}}_0)$ and $S^{\hat{S}_{\mathbf{r}},\hat{n}}_{\mathrm{F}} = -\mathrm{log}(I^{\hat{S}_{\mathbf{r}}}_0 + I^{\hat n}_0)$ respectively. Here, $\hat{S}_{\mathbf{r}}$ is chosen to minimize the coherence and diagonal terms in Eq.~(\ref{eqn:EE_MQC}) (Supplementary Methods).
 (b) Long-time spin-phonon RE $S_2(\hat{\rho}_{\mathrm{ph}})$ as a function of transverse field. To remove finite-size effects and residual oscillations we plot a time-averaged value $\overline{S_2(\hat{\rho}_{\mathrm{ph}})}$ for $4~\mathrm{ms}\leq t \leq 12$~ms (FOTOC quantities are averaged identically). The regular and chaotic dynamics for the initial state $\vert \Psi^c_0\rangle$ are clearly delineated: $\overline{S_2(\hat{\rho}_{\mathrm{ph}})} \approx 0$ for $B > B_{\mathrm{c}}$ and $\overline{S_2(\hat{\rho}_{\mathrm{ph}})} > 0$ for $B < B_{\mathrm{c}}$ respectively. Error bars indicate standard deviation of temporal fluctuations. In the inset we plot the same FOTOC quantities but including decoherence due to single-particle dephasing at the rate $\Gamma = 60$~s$^{-1}$. The coherent parameters $g$, $B$ and $\delta$ are enhanced by a factor of 16 compared to the main panel, as per Ref.~\cite{FossFeigPA}.
 (c) Time-averaged distribution functions (markers) for spin-projection $P(M_z)$ and phonon occupation $P(n)$ ($6~\mathrm{ms} \leq t \leq 12~\mathrm{ms}$). We compare to the distribution of the diagonal ensemble (bars, see Methods).
 (d) Bipartite RE $S_2(\hat{\rho}_{L_A})$ (black markers) as a function of partition size ${L_A}$ of the spins, averaged over same time window as (c). For comparison, we plot the RE of a thermal canonical ensemble with corresponding temperature $T$ fixed by the energy of the initial state $\vert \Psi_0^c \rangle$, $S^{\mathrm{therm}}_2$ and the RE of the diagonal ensemble (see Methods). Volume-law behaviour of the RE is replicated by the FOTOC quantity (blue markers). Note that the dimension of the spin Hilbert space scales linearly with $L_A$. Shaded regions indicate standard deviation of temporal fluctuations.  Data for (a)-(d) is obtained for $N=40$, with $g$ and $\delta$ identical to calculations of Fig.~\ref{fig:phase}. For (c) and (d) we choose $B/(2\pi) = 0.7$~kHz ($B/B_{\mathrm{c}} = 0.2$). Source data are provided as a Source Data file.}
 \label{fig:ThermErg}
\end{figure}

\noindent{\it \bf Thermal and diagonal ensembles}
The canonical thermal ensemble, used in Fig.~\ref{fig:ThermErg}, is defined by the density matrix $\hat{\rho}_{\mathrm{therm}} = e^{-\beta\hat{H}_D}/\mathrm{Tr}[e^{-\beta\hat{H}}]$, which is characterized by the inverse temperature $\beta = 1/(k_B T)$. This inverse temperature is chosen such that energy of the ensemble is matched to that of the initial state of the dynamics, $\langle E \rangle_{\mathrm{therm}} \equiv \mathrm{Tr}[\hat{H}_D \hat{\rho}_{\mathrm{therm}}]= \langle \Psi_0 \vert \hat{H}_D \vert \Psi_0 \rangle$. The RE for bipartitions of the thermal ensemble is then obtained via the definition $S_2^{\mathrm{therm}} \equiv -\mathrm{log}\left(\mathrm{Tr}[(\hat{\rho}^{\mathrm{therm}}_{L_A})^2]\right)$ where $\hat{\rho}^{\mathrm{therm}}_{L_A} = \mathrm{Tr}_{\mathrm{ph},N-L_A}(\hat{\rho}_{\mathrm{therm}})$ is the reduced density matrix obtained after tracing out the phonon degree of freedom and the remaining $N-L_A$ spins. 

A related concept is the diagonal ensemble $\hat{\rho}_D$ \cite{DAlessio2016,Rigol2008}, which generically describes the (time-averaged) observables of a quantum system which has relaxed at long times. The ensemble is defined as the mixed state $\hat{\rho}_D \equiv \sum_{E_n} \vert c_{E_n} \vert^2 \vert E_n \rangle \langle E_n \vert$ where $c_{E_n} \equiv \langle \Psi_0 \vert E_n \rangle$ and $\vert E_n \rangle$ are the eigenstates of the Hamiltonian $\hat{H}_D$ with asociated eigenvalue $E_n$. We use this diagonal ensemble as a comparison to the time-averaged distribution functions $P(M_Z)$ and $P(n)$ in Fig.~\ref{fig:ThermErg}

\noindent {\bf Data Availability:} The source data underlying Figs.~2-4 of the main text are provided as a source data file. Additional numerical data and computer codes used in this study are available from the corresponding author upon request.

\bibliography{library}

\noindent {\bf Acknowledgements:} We thank  A. Kaufman and R.  Nandkishore for fruitful discussions. This work is supported  by the Air Force Office of Scientific Research grants   FA9550-18-1-0319  and its Multidisciplinary University Research Initiative grant(MURI), by  the Defense Advanced Research Projects Agency (DARPA) and Army Research Office grant W911NF-16-1-0576, the National Science Foundation grant PHY-1820885, JILA-NSF grant PFC-173400, and the National Institute of Standards and Technology.

\noindent {\bf Additional Note:} Upon completion of this manuscript we became aware of the recent preprints \cite{1808.02038,1807.10292}, which present numerical and analytic investigation of OTOCs in the Dicke model.

\noindent {\bf Author Contributions:} The calculations were performed by R. L-S. and A. S-N. All authors participated in the conception of the project, analysis of the results and preparation of the manuscript.

\renewcommand{\baselinestretch}{1.0}

\clearpage

{\it \bf Supplemental Material}

\noindent{\it \bf Supplementary Methods} 
In the main text we leave unspecified the exact form of the time-dependent spin-rotation $\hat{S}_{\mathbf{r}}$ from which the MQC $I^{\hat{S}_{\mathbf{r}}}_0(t)$ is obtained in, e.g. Fig.~4 of the main text, although we have argued that any generic rotation should yield a good approximation to the Renyi entropy. However, for the still relatively small systems we consider it is clear that there will exist an `optimal' rotation choice for which our arguments will give the best quantitative correspondence, i.e. a basis in which the state appears the `most scrambled'. By this we mean that when the density matrix is written in the basis of the optimal rotation $\hat{S}_{\mathbf{r}}$, the state has a broad spin probability distribution $P(M_{\mathbf{r}})$ and there are a large number of essentially random off-diagonal coherences which respectively lead to a suppression of $D^{\hat{S}_{\mathbf{r}},\hat{n}}_{\mathrm{diag}}$ and $C^{\hat{S}_{\mathbf{r}},\hat{n}}_{\mathrm{off}}$.

The (time-dependent) choice of $\hat{S}_{\mathbf{r}}$ which guarantees this is not neccesarily clear \emph{a priori}. However, in Supplementary Fig.~\ref{fig:OptOTOC} we demonstrate that: i) For sufficiently scrambled states of the Dicke model, any choice of $\hat{S}_{\mathbf{r}}$ gives a qualitatively robust correspondence to the entanglement entropy, and ii) an educated guess for $\hat{S}_{\mathbf{r}}$ can be made without rigourously optimising the FOTOCs over all rotation axes, but rather by searching for the maximum variance $\mathrm{var}(\hat{S}_{\mathbf{r}})$ after time $t$ (i.e., the first half of the many-body echo sequence). The latter can be understood to be a crude measure of how delocalized the distribution $P(M_{\mathbf{r}})$ is. We do reiterate, however, that the results plotted in Supplementary Fig.~\ref{fig:OptOTOC} clearly show that any rotation axis does always track the qualitative and indeed quantitative structure of the full Renyi entropy, and the optimisation is thus only fine-tuning.

\begin{figure}
    \includegraphics[width=8cm]{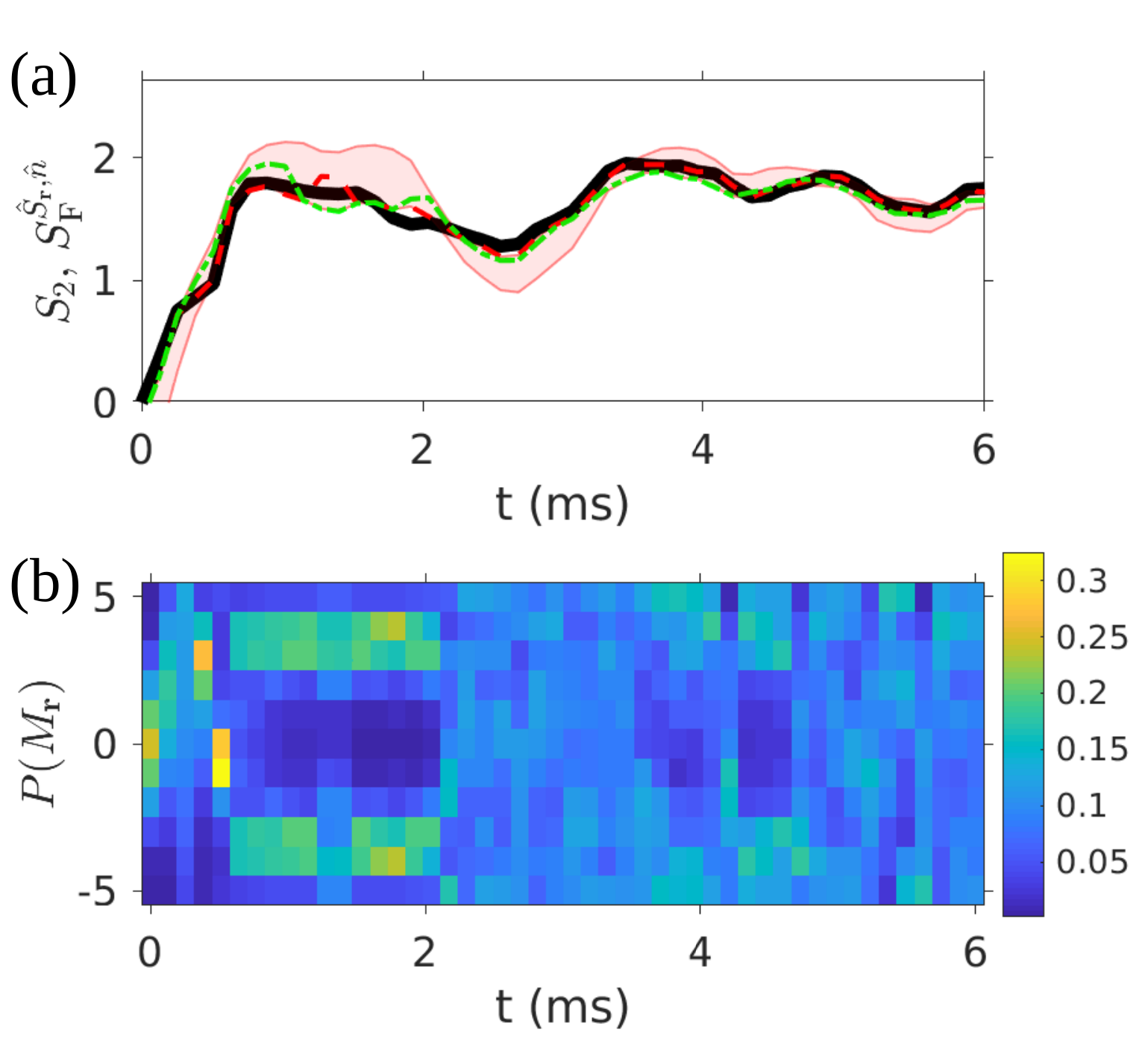}
    \caption{Optimisation of FOTOC rotation axis. (a) Typical evolution of spin-phonon entanglement $\hat{S}_2(\hat{\rho}_{\mathrm{ph}})$ (black solid line) in the chaotic phase. We find excellent agreement with the FOTOC quantity $S^{\hat{S}_{\mathbf{r}},\hat{n}}_{\mathrm{F}}$ for both rigorous optimisation of $\hat{S}_{\mathbf{r}}$ (red dashed line), optimisation via maximisation of $\mathrm{var}( \hat{S}_{\mathbf{r}})$ (green dot-dashed line). Shaded red regions represent full range of possible $S^{\hat{S}_{\mathbf{r}},\hat{n}}_{\mathrm{F}}$ for any rotation axis.
    (b) Spin probability distribution $P(M_{\mathbf{r}})$ for optimal rotation $\hat{S}_{\mathbf{r}}$ [red dashed line in (a)]. For $t\gtrsim 2$~ms the distribution becomes delocalized, consistent with the arguments in text.
    Data for both panels is for $B/B_c \approx 0.1$, $N=10$ and all other parameters as per Fig.~4 of the main text. Source data are provided as a Source Data file.}
    \label{fig:OptOTOC}
\end{figure}

% \vspace{0.3in}

\noindent{\it \bf Supplementary Note 1} 
In the main text we present data for $\lambda_{\mathrm{Q}}$ and $\lambda_{\mathrm{L}}$ in Supplementary Fig.~3 obtained from sample initial states $\vert\Psi_0^c\rangle=\vert -N/2\rangle_x \otimes \vert 0\rangle$ and $\vert\Psi^y_0\rangle=\vert -N/2\rangle_y \otimes \vert 0\rangle$ as a function of $B/B_c$ and the FOTOC $\mathcal{F}_X(t)$. The results plotted in Supplementary Fig.~3 validate our predicted relation $\lambda_{\mathrm{Q}} \simeq 2\lambda_{\mathrm{L}}$.

Here, we elaborate on this data in two ways. First, we show the generic growth of the FOTOC for not only the exemplary state $\vert\Psi_0^c\rangle$ (also shown in Fig.~3 of the main text) but also $\vert\Psi^y_0\rangle$. These results are plotted in Supplementary Fig.~\ref{fig:ExampleExp} and demonstrate that generically, we do not observe perfect exponential growth, as for the state $\vert\Psi_0^c\rangle$, but rather an oscillatory function which grows with an exponential trend. We plot against $\sim e^{2\lambda_{\mathrm{L}} t}$ for comparison. The exemplary nature of the exponential growth for $\vert\Psi_0^c\rangle$ is attributable to it being an unstable fixed point of the classical phase-space with $\langle \hat{X} \rangle = \overline{\alpha_R} = 0$.

Second, we also demonstrate the broad validity of our relation between the quantum and classical exponents for other FOTOCs/operators. Specifically, we plot the growth of $\mathcal{F}_{S_y}(t)$ and $\mathcal{F}_{n}(t)$. For $\mathcal{F}_{S_y}(t)$, we observe exponential growth similar to the results of $\mathcal{F}_{X}(t)$ such that $\mathcal{F}_{S_y}(t) \sim e^{\lambda_{\mathrm{Q}} t}$ and $\lambda_Q \approx 2\lambda_{\mathrm{L}}$ as previous. However, the results for $\mathcal{F}_{n}(t)$ require some further explanation. Specifically, $\hat{n} = \hat{a}^{\dagger}\hat{a}$ is nonlinear in the classical variables, i.e. $\hat{n} \to n \equiv \alpha_R^2 + \alpha_I^2$. Whilst in the classical model the seperation of trajectories measured in terms of the nonlinear variable $n$ will still grow exponentially, they do so with a different classical exponent: $\vert n_1(t) - n_2(t) \vert \approx \vert n_1(0) - n_2(0) \vert e^{\lambda_{\mathrm{c}} t}$. As $n$ is not an invertible transformation of the co-ordinates $\alpha_R$ and $\alpha_I$ then this classical exponent $\lambda_{\mathrm{c}}$ is not neccesarily identical to the Lyapunov exponent $\lambda_{\mathrm{L}}$ determined from the distance in terms of the natural phase-space variables. Indeed, for $n$ we typically observe $\lambda_{\mathrm{c}} \approx 2 \lambda_{\mathrm{L}}$ (see Supplementary Fig.~\ref{fig:lambdaLvsC}). The key consequence of this subtlety is that while we still observe exponential growth of the FOTOC $\mathcal{F}_{n}(t) \sim e^{\lambda^{\prime}_{\mathrm{Q}} t}$, we have that $\lambda^{\prime}_{\mathrm{Q}} \approx 2\lambda_{\mathrm{c}}$ where $\lambda^{\prime}_{\mathrm{Q}}$ is thus not neccesarily identical to $\lambda_Q$ obtained from $\mathcal{F}_X(t)$ (or other FOTOCs formed from linear combinations of the phase-space co-ordinates). Thus, a more general statement of our finding relating the quantum and Lyapunov exponents in the main text is:
the quantum exponent of an exponentially growing FOTOC is approximately twice that of the appropriately defined classical exponent, $\lambda_{\mathrm{Q}} = 2\lambda_{\mathrm{c}}$. In the case where the FOTOC operator $\hat{G}$ corresponds to a linear combination of the phase-space co-ordinates this reduces to $\lambda_{\mathrm{Q}} = 2\lambda_{\mathrm{L}}$ as $\lambda_{\mathrm{c}} \equiv \lambda_{\mathrm{L}}$. This result is reflected in the results plotted in Supplementary Fig.~\ref{fig:ExampleExp}.

\begin{figure}[h]
    \includegraphics[width=8cm]{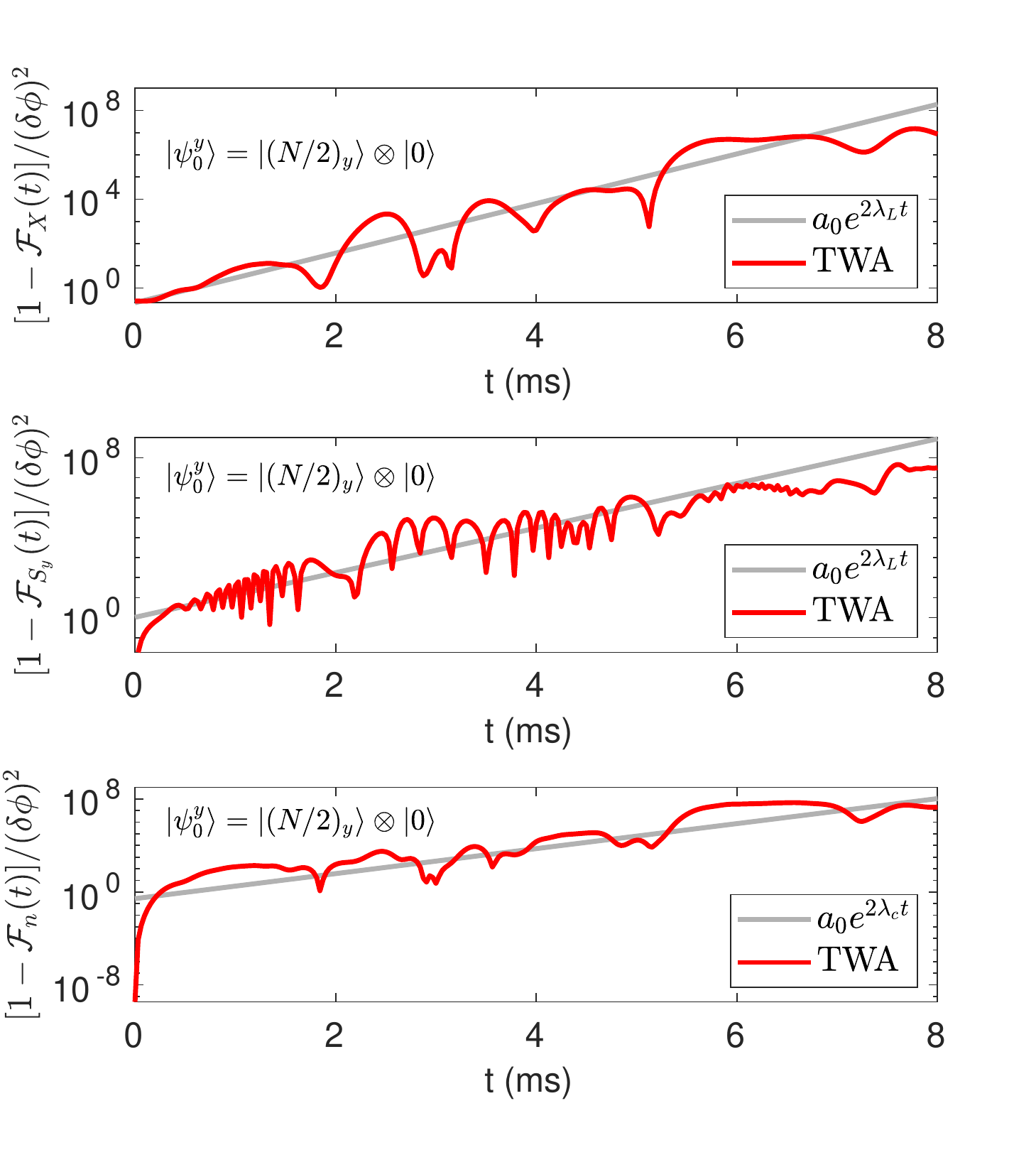}
    \includegraphics[width=8cm]{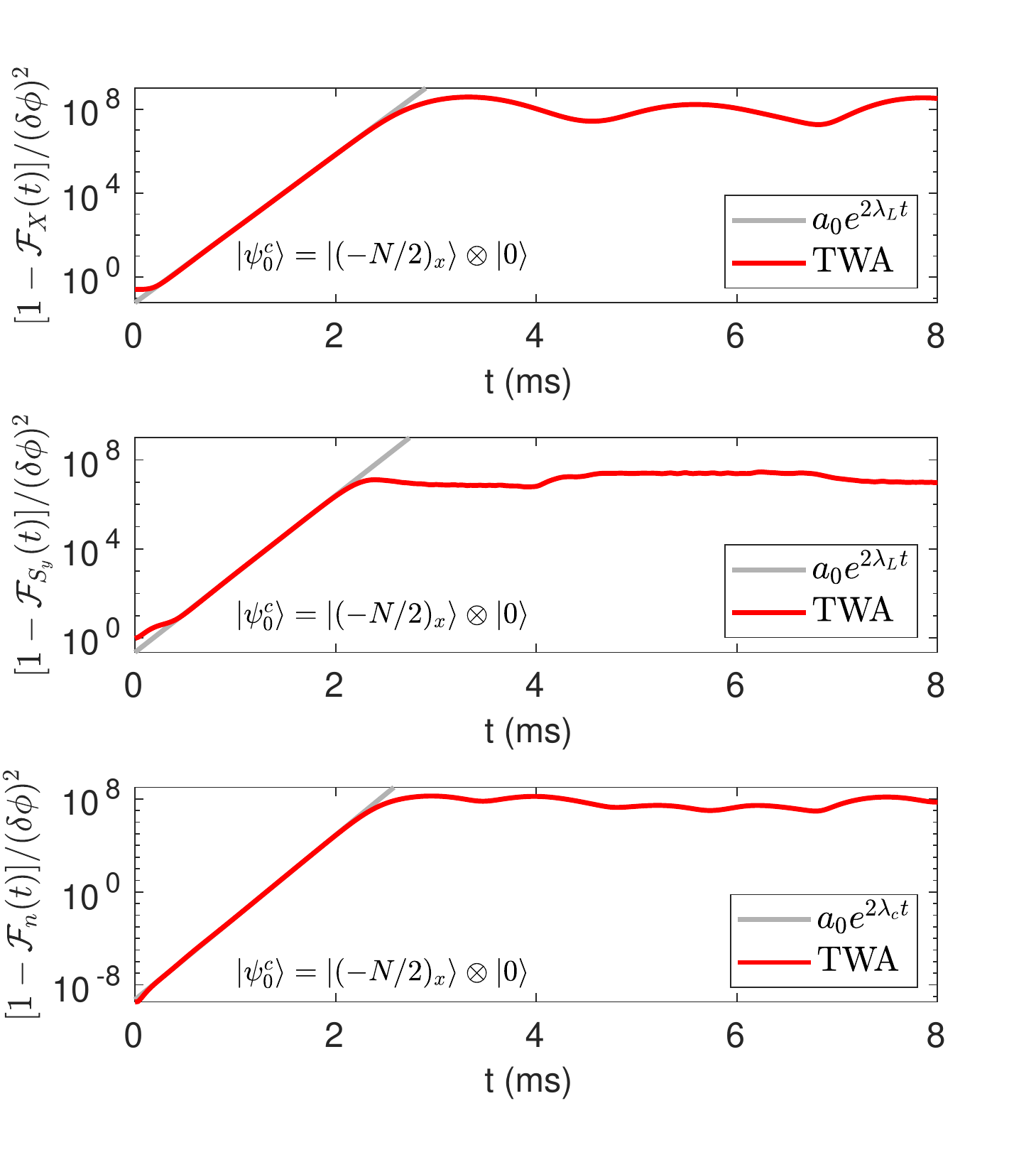}
    \caption{Exponential growth of quantum variances $\mathrm{var}(\hat{X}) \approx [1 - \mathcal{F}_{X}(t)]/(\delta\phi)^2$, $\mathrm{var}(\hat{S}_y) \approx [1 - \mathcal{F}_{S_y}(t)]/(\delta\phi)^2$ and $\mathrm{var}(\hat{n}) \approx [1 - \mathcal{F}_{n}(t)]/(\delta\phi)^2$ (red lines) for $N=10^8$ from truncated Wigner calculations (assuming $\delta\phi \ll 1/N$). We give examples for $\vert\Psi_0^c\rangle=\vert (-N/2)_x\rangle \otimes \vert 0\rangle$ (right) and $\vert\Psi^y_0\rangle=\vert (N/2)_y\rangle \otimes \vert 0\rangle$ (left). Grey lines indicate a comparison to $a_0 e^{2\lambda_{\mathrm{L,c}} t}$, with $a_0$ fitted to FOTOC data. Other parameters are same as calculations in Fig.~3a of the main text. Source data are provided as a Source Data file.}
    \label{fig:ExampleExp}
\end{figure}

\begin{figure}[h]
    \includegraphics[width=8cm]{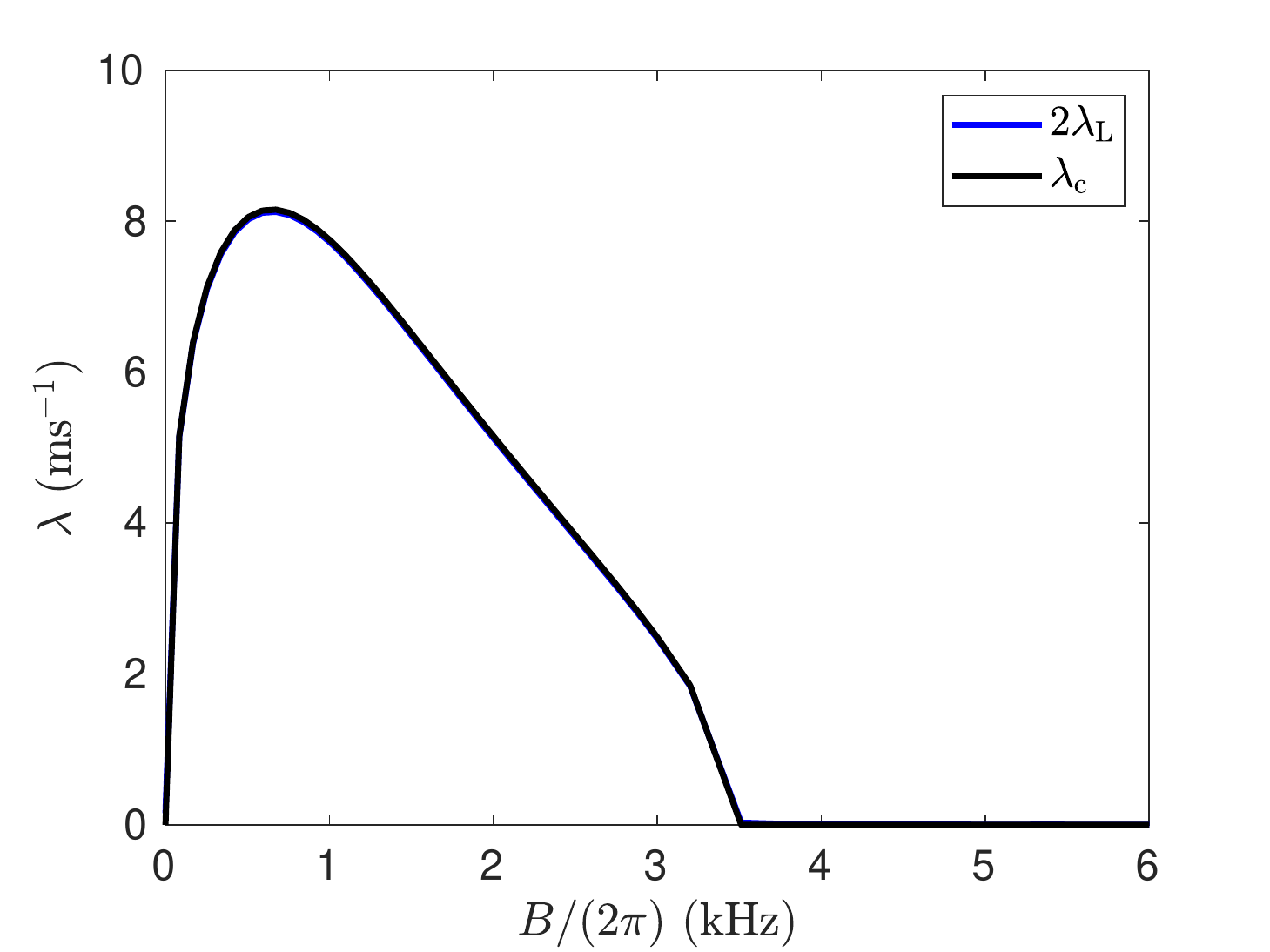}
    \caption{Comparison of classical Lyapunov exponent $\lambda_{\mathrm{L}}$ and $\lambda_{\mathrm{c}}$ as a function of transverse field strength $B/B_{\mathrm{c}}$. The latter is obtained by solution of the classical equations of motion (see Methods section of main text) and the definition $\vert n_1(t) - n_2(t) \vert \approx \vert n_1(0) - n_2(0) \vert e^{\lambda_{\mathrm{c}} t}$ where the subscript labels two trajectories which are initially close in phase-space. The exponents are calculated for the initial condition $\tilde{\bf x} = (-N/2,0,0,0,0)$ corresponding to the quantum state $\vert \Psi_0^c \rangle$. Parameters are as per Figs. 2 and 3 of the main text. Source data are provided as a Source Data file.}
    \label{fig:lambdaLvsC}
\end{figure}

% \vspace{0.3in}

\noindent{\it \bf Supplementary Note 2} 
Here we outline a generalization of FOTOCs and the related multiple quantum intensities for a generic system. In particular, we consider how FOTOCs might be implemented for spin models which are not collective. We highlight, however, that our analysis can be straightforwardly generalized to other systems, such as bosonic models involving many modes (which appear in, e.g., quantum gas microscope experiments).  Our analysis is closely related to that presented in pages 7-8 of the manuscript and the methods.

Let us consider a generic system $\mathcal{S}$ of $N$ spin-$1/2$s, which can be arbitrarily split into two subsystems $\mathcal{A}$ and $\mathcal{B}$ containing $N_{\mathcal{A}}$ and $N_{\mathcal{B}}$ spins respectively. The system evolves under an arbitrary (but non-collective) Hamiltonian $\hat{H}$. We can write generic pure states of the system in terms of the basis $\{\vert \vec{a} \rangle\}$ and $\{\vert \vec{b} \rangle\}$, which are defined to be the tensor product of single particle states such that $\vert \vec{a} \rangle \equiv \bigotimes_j \vert a^{\mathbf{r}_j}_j \rangle$ (and similarly for $\vert\vec{b}\rangle$) where $\hat{\sigma}^{\mathbf{r}_j}_j\vert a^{\mathbf{r}_j}_j \rangle \equiv \pm \vert a^{\mathbf{r}_j}_j \rangle$ are eigenstates of the spin-projection along an arbitrary direction $\mathbf{r}_j$ on the spin-$1/2$ Bloch sphere for the $j$th qubit in $\mathcal{A}$. We emphasize that most generically the spin-projection is defined along an independent direction $\mathbf{r}_j$ for each of the qubits.

It is then straightforward to decompose the density matrix (similar to page 8 on the main text) in this basis as:
\begin{equation}
    \hat{\rho} = \sum_{\substack{\vec{a},\vec{a}^{\prime} \\ \vec{b}, \vec{b}^{\prime}}} \varrho^{\vec{a},\vec{a}^{\prime}}_{\vec{b},\vec{b}^{\prime}} \vert \vec{a} \rangle \langle \vec{a}^{\prime} \vert \otimes \vert \vec{b} \rangle \langle \vec{b}^{\prime} \vert .
\end{equation}
Similar to the collective model, we divide the density matrix into blocks of (single-particle) coherences $\hat{\rho} \equiv \sum_{\vec{M}} \hat{\rho}^{\{\hat{G}\}}_{\vec{M}}$ with respect to a set of single-particle operators $\{\hat{G}\}$. Specifically, we will focus on the case where $\{\hat{G}\} = \{\hat{\sigma}^{\mathbf{r}_j}_j| j \in \mathcal{A}\}$ or $\{\hat{G}\} = \{\hat{\sigma}^{\mathbf{r}_j}_j| j \in \mathcal{B}\}$ separately. For the former, we then define each block as
% \begin{multline}
    % \hat{\rho}^{\{\hat{G}\}}_{\vec{M}} = \sum_{ \substack{ \vec{n} \in \mathcal{A} \\  \vec{m} \in \mathcal{A}_c , \vec{m}^{\prime} \in \mathcal{A}_c } } \varrho^{n^{\mathbf{r}_1}_1, n^{\mathbf{r}_2}_2, ..., n^{\mathbf{r}_1}_1 + M_1, n^{\mathbf{r}_2}_2 + M_2, ...  }_{ m^{\mathbf{r}_{N_{\mathcal{A}+1}}}_{N_{\mathcal{A}+1}}, m^{\mathbf{r}_{N_{\mathcal{A}+2}}}_{N_{\mathcal{A}+2}} } \\
    % \times \left[ \vert n^{\mathbf{r}_1}_1 \rangle \langle n^{\mathbf{r}_1}_1 + M_1 \vert \otimes \vert n^{\mathbf{r}_2}_2 \rangle \langle n^{\mathbf{r}_2}_2 + M_2 \vert \otimes ... \otimes \vert m^{\mathbf{r}_{N_{\mathcal{A}+1}}}_{N_{\mathcal{A}+1}} \rangle \langle m^{\mathbf{r}_{N_{\mathcal{A}+1}}}_{N_{\mathcal{A}+1}} \vert \otimes \vert m^{\mathbf{r}_{N_{\mathcal{A}+2}}}_{N_{\mathcal{A}+2}} \rangle \langle m^{\mathbf{r}_{N_{\mathcal{A}+2}}}_{N_{\mathcal{A}+2}} \vert \otimes ... \right]
% \end{multline}
\begin{multline}
    \hat{\rho}^{\{\hat{G}\}}_{\vec{M}} = \sum_{ \substack{ \vec{a} \in \mathcal{A} \\  \vec{b} \in \mathcal{B} , \vec{b}^{\prime} \in \mathcal{B} } }
    \varrho^{a^{\mathbf{r}_1}_1, a^{\mathbf{r}_2}_2, ..., a^{\mathbf{r}_1}_1 + M_1, a^{\mathbf{r}_2}_2 + M_2, ...  }_{ b^{\mathbf{r}_{1}}_{1}, b^{\mathbf{r}_{2}}_{2}, ..., b^{\mathbf{r}_{1}}_{1}, b^{\mathbf{r}_{2}}_{2}, ... }
    \\
    \times \Big[ \vert a^{\mathbf{r}_1}_1 \rangle \langle a^{\mathbf{r}_1}_1 + M_1 \vert \otimes \vert a^{\mathbf{r}_2}_2 \rangle \langle a^{\mathbf{r}_2}_2 + M_2 \vert \otimes ... \otimes \vert b^{\mathbf{r}_{1}}_{1} \rangle \langle b^{\mathbf{r}_{1}}_{1} \vert \otimes \vert b^{\mathbf{r}_{2}}_{2} \rangle \langle b^{\mathbf{r}_{2}}_{2} \vert \otimes ... \Big] . \label{eqn:GeneralizedRhoM}
\end{multline}
Whilst this expression may look daunting, one can still define a set of generalized multiple quantum intensities $I^{\{\hat{G}\}}_{\vec{M}} \equiv \mathrm{Tr}\left[ \hat{\rho}^{\{\hat{G}\}}_{\vec{M}} \hat{\rho}^{\{\hat{G}\}}_{-\vec{M}} \right]$, which  are related to the generalized FOTOC $F_{\{\hat{G}\}}(t,\phi_1,\phi_2,...) = \sum_{\vec{M}} I^{\{\hat{G}\}}_{\vec{M}} e^{-i\sum_j M_j\phi_j}$. This FOTOC is experimentally implemented by applying single-qubit rotations on the spins contained in $\mathcal{A}$ (alternately, $\mathcal{B}$) such that $\hat{W}_{\{\hat{G}\}} \equiv \bigotimes_{j\in \mathcal{A}} e^{i\phi_j\hat{\sigma}^{\mathbf{r}_j}_j}$. Here, the $j$th qubit is rotated about an independently chosen axis $\mathbf{r}_j$ by an angle $\phi_j$.

Again, of most interest to us will be the $0$-th multiple quantum intensity $I^{\{\hat{G}\}}_{\vec{0}} \equiv \mathrm{Tr}\left[ \right(\hat{\rho}^{\{\hat{G}\}}_{\vec{0}} \left)^2 \right]$ [i.e. $M_j = 0$ for all $j$ in Eq.~(\ref{eqn:GeneralizedRhoM})]. This is accessed experimentally by performing a full set of OTOCs for a rangle of angles $\{\phi_j\}$, such that $I^{\{\hat{G}\}}_{\vec{0}} \propto \sum_{\{\phi_j\}} F_{\{\hat{G}\}}(t,\phi_1,\phi_2,...)$. Similar to the main text, we can use this quantity to write the purity of the reduced density matrix in $\mathcal{A}$, $\hat{\rho}_{\mathcal{A}}$, as:
\begin{equation}
    \mathrm{Tr}\left[ \hat{\rho}_{\mathcal{A}}^2 \right] \equiv I^{\{\hat{G}\}_{\mathcal{A}}}_{\vec{0}} + I^{\{\hat{G}\}_{\mathcal{B}}}_{\vec{0}} - D^{\{\hat{G}\}_{\mathcal{A}}, \{\hat{G}\}_{\mathcal{B}}}_{\mathrm{diag}} + C^{\{\hat{G}\}_{\mathcal{A}}, \{\hat{G}\}_{\mathcal{B}}}_{\mathrm{off}}
\end{equation}
where $\{\hat{G}\}_{\mathcal{A}} \equiv \{\hat{\sigma}^{\mathbf{r}_j}_j| j \in \mathcal{A}\}$ and similar for $\{\hat{G}\}_{\mathcal{B}}$. The latter terms are given by
\begin{equation}
    D^{\{\hat{G}\}_{\mathcal{A}}, \{\hat{G}\}_{\mathcal{B}}}_{\mathrm{diag}} = \sum_{\vec{a},\vec{b}} \left[ \varrho^{\vec{a},\vec{a}}_{\vec{b},\vec{b}} \right]^2 ,
\end{equation}
and
\begin{equation}
    C^{\{\hat{G}\}_{\mathcal{A}}, \{\hat{G}\}_{\mathcal{B}}}_{\mathrm{off}} = \sum_{ \substack{\vec{a}\neq\vec{a}^{\prime} \\ \vec{b}\neq\vec{b}^{\prime} } }  \varrho^{\vec{a},\vec{a}^{\prime}}_{\vec{b},\vec{b}} \varrho^{\vec{a}^{\prime},\vec{a}}_{\vec{b}^{\prime},\vec{b}^{\prime}} .
\end{equation}
Following the main text and methods, identical arguments can be made that the contributions from $D^{\{\hat{G}\}_{\mathcal{A}}, \{\hat{G}\}_{\mathcal{B}}}_{\mathrm{diag}}$ and $C^{\{\hat{G}\}_{\mathcal{A}}, \{\hat{G}\}_{\mathcal{B}}}_{\mathrm{off}}$ vanish in certain scenarios for a generic choice of operators $\{\hat{G}\}_{\mathcal{A},\mathcal{B}}$. In particular, the machinery of canonical pure thermal quantum (cTPQ) states can be used in the case of non-integrable systems to verify that
$C^{\{\hat{G}\}_{\mathcal{A}}, \{\hat{G}\}_{\mathcal{B}}}_{\mathrm{off}} \to 0$ for sufficiently large systems and $D^{\{\hat{G}\}_{\mathcal{A}}, \{\hat{G}\}_{\mathcal{B}}}_{\mathrm{diag}} \ll I^{\{\hat{G}\}_{\mathcal{A}}}_{\vec{0}} + I^{\{\hat{G}\}_{\mathcal{B}}}_{\vec{0}}$ after short times.

The requirement of single-qubit rotations is similar in spirit to the protocol of random measurements proposed in Ref.~[24], although the exact connection between the proposed schemes is an open question. We point out that alternatively one can also extract the required multiple quantum intensities by direct measurement of joint probability distribution functions in the chosen bases:
\begin{eqnarray}
 I^{\{\hat{G}\}_{\mathcal{A}}}_{\vec{0}} \equiv \sum_{a \in \mathcal{A}} P(a^{\mathbf{r}_1}_1,a^{\mathbf{r}_2}_2,...)^2 , \label{eqn:I0A_single} \\
\end{eqnarray}
and similarly for $I^{\{\hat{G}\}_{\mathcal{A}_c}}_{\vec{0}}$. This form might be useful in, e.g., small chains of trapped ions and quantum gas microscope experiments (using a bosonic occupation basis), wherein one can measure such joint distributions relatively efficiently.

Lastly, we note that one can demonstrate that collective rotations of the spins in $\mathcal{A}$ and the associated $I_0^{\hat{S}^\mathcal{A}_{\mathbf{r}}}$ can be related to the equivalent $I^{\{\hat{G}\}_{\mathcal{A}}}_{\vec{0}}$ obtained via a uniform set of rotations with $\mathbf{r}_j = \mathbf{r}$. In particular, we have that the terms of $I^{\{\hat{G}\}_{\mathcal{A}}}_{\vec{0}}$ are contained within an appropriate expansion of $I_0^{\hat{S}^\mathcal{A}_{\mathbf{r}}}$ in the single-particle basis. This indicates that even for systems which span beyond the fully symmetric Dicke basis, collective rotations may allow us to gain some insight into the Renyi entropy. This is, however, an open question and currently under investigation.

\end{document}